\documentclass[twocolumn,prb,preprintnumbers,amsmath]{revtex4}
\usepackage{epsfig}
\usepackage{lscape}
\usepackage{graphicx}
\usepackage{graphics}
\usepackage{bm}
\usepackage[usenames]{color}

\begin{document}
\preprint{LA-UR-07-0685}

\title{Environment Dependent Charge Potential for Water}

\author{Krishna Muralidharan}
\affiliation{Department of Physics and Astronomy, University of
New Mexico, Albuquerque, New Mexico 87131 and \\
Quantum Theory Project, University of Florida, Gainesville, FL 32601}

\author{Steven~M.~Valone}
\affiliation{Materials Science and Technology Division, Los Alamos
National Laboratory,
Los Alamos, New Mexico 87545 and \\
Department of Physics and Astronomy, University of New Mexico,
Albuquerque, New Mexico 87131}

\author{Susan~R.~Atlas}
\email{susie@sapphire.phys.unm.edu}
\affiliation{Center for Advanced Studies and
Department of Physics and Astronomy, \\ University of New Mexico,
Albuquerque, New Mexico 87131
\vspace*{0.05in}}

\date{\today}

\begin{abstract}
We present a new interatomic potential for water captured in a
charge-transfer embedded atom method (EAM) framework. The potential
accounts for explicit, dynamical charge transfer in atoms as a
function of the local chemical environment. As an initial test of
the charge-transfer EAM approach for a molecular system, we have
constructed a relatively simple version of the potential and
examined its ability to model the energetics of small water
clusters. The excellent agreement between our results and current
experimental and higher-level quantum computational data signifies a successful
first step towards developing a unified charge-transfer potential
capable of accurately describing the polymorphs, dynamics, and
complex thermodynamic behavior of water.
\end{abstract}
\maketitle

\section{Introduction}

\medskip

Developing accurate interatomic potentials is essential for modeling
the atomistic behavior of materials in diverse chemical and physical
environments. Broadly speaking, potentials can be classified as
empirical or semi-empirical. Empirical potentials---ranging from the
relatively simple Lennard-Jones and Buckingham potentials to more
elaborate force fields such as CHARMM\cite{CHARMM} or ReaxFF\cite{ReaxFF}---are
typically parameterized to represent a set of specific structural or
thermodynamic properties of a given molecular system or material.
Consequently they may have limited success in representing other
non-parameterized properties, or materials whose atomic or molecular
constituents lie outside the parametrization set. Systems where
charge transfer effects are important present a particular challenge
to empirical approaches. The simplest models incorporate fixed
formal atomic charges with distance-dependent charge transfer
switching functions.\cite{EMP-CT1} Others define a charge-dependent
functional form,\cite{EMP-CT2} often a simple quadratic as in the
ES+ method,\cite{StrM} and adjust the charges via chemical potential
equalization. On the other hand, semi-empirical potentials guided by
quantum mechanics (QM)---for example, the embedded-atom
(EAM)\cite{EAM,EAM2,EAM-rev} and modified embedded-atom methods
(MEAM),\cite{MEAM} tight-binding (TB) theory,\cite{TB}
SCC-DFTB/CHARMM,\cite{SCC-DFTB} diatomics-in-molecules\cite{DIM} and
empirical valence bond (EVB) approaches\cite{EVB,EVB-Voth}---depend
on potential parameters derived from {\it ab initio} calculations or
experimental data.  The success of these semi-empirical approaches
hinges upon the ability of the model to assimilate relevant QM and
experimental information within a functional form that depends on a
relatively small set of parameters, and can be translated readily
into computer code for efficient application to large-scale
simulation systems. In the case of the TB and EVB
approaches, the parameterizations generally require the
specification of a carefully tailored set of basis wavefunctions,
the estimation of corresponding overlap integrals, and on-the-fly
Hamiltonian diagonalizations.
These explicit QM steps significantly complicate the construction of
the potentials, thus limiting the size and chemical diversity of the
systems to which they can be applied. Similarly, SCC-DFTB/CHARMM and
other QM/MM methods\cite{QMMM} require the definition of appropriate
auxiliary conditions in order to handle boundary-matching, charge
polarization, and long-range electrostatic interactions between the
quantum and classical (molecular mechanics) regions of the system.
Semi-empirical potentials such as EAM and MEAM do not involve
explicit QM components, but in their present form cannot account for
non-perturbative changes in the charge states of atoms.
Consequently, they are not expected to accurately model such
important biophysical and materials problems as polar systems,
electron transport, defect-driven charge polarization, fluctuating
valence systems, complex oxides, and reactive dynamics.

In this study, we present the first implementation of a novel
charge-transfer embedded atom (CT-EAM) potential aimed at addressing
the issues outlined above, and apply it to the structure and
energetics of netural water clusters $({\rm H}_2{\rm O})_n$,
$n=2,\ldots ,20$. Importantly, the new potential incorporates
quantum mechanical information in the spirit of TB, EVB, and related
approaches, while preserving the intuitive features, ease of
parametrization, and extensibility of the EAM. The potential is
based on a multiscale framework recently described by two of the
authors.\cite{vat06,atv06}  This framework is formally based in
density functional theory (DFT),\cite{hok,kos} and couples the
electronic and atomistic length scales within a self-consistent
classical potential.  A key feature of the potential is its
dependence on the redistributed atomic electron densities---and by
extension, charge transfer---which vary with the instantaneous
configuration of the atoms within the molecule or material. The
parameterized charge distributions are derived from \textit{ab
initio} calculations. This `atom-in-molecule' perspective and
associated effective charge are at the heart of the CT-EAM model
framework. A second important aspect is the imposition of
self-consistency between the atomic electron densities appearing in
the two physically distinct---embedding and
electrostatic---components of the model. This requirement is
intrinsic to the CT-EAM theory, and is in contrast to other charge
transfer models, including some based on the EAM, where different
functional forms are assigned to nominally identical electron
densities.

For this initial implementation, we focus on water as a paradigmatic
small molecule system of immense practical importance to
biomolecular and materials applications. Water also represents an
extremely challenging test system for any classical potential due to
its strong polar features arising from underlying charge transfer and
charge polarization, and associated many-body effects.\cite{Stone07}

In the following section we briefly review the extensive literature
on classical potentials for water, with emphasis on previous
approaches to the treatment of charge transfer. We then review the
EAM method and describe previous attempts to adapt the model to the
study of charge-transfer systems. This is followed by a summary of
the key features of the recently proposed CT-EAM as implemented in
the present work. In Section III, we present our potential
parameterization, and in Section IV, the results of the model for
various water clusters. The paper concludes with a summary and
discussion of future work.

\section{Background}

\subsection{Water potentials}

Understanding the thermodynamic and structural properties of water
is crucial to modeling many biological, chemical and physical
phenomena. Despite its relevance and importance, there are still
unanswered questions regarding the properties of water polymorphs
and their exact roles in solution chemistry as well as in biological
processes. Developing an accurate model capable of simultaneously
describing the gas phase, liquid, and solid state properties of
water has presented enormous challenges. Ultimately, a complete
model should be capable of describing diverse phenomena such as ion
solvation,\cite{MEVB} electro-,\cite{SEC55} photo-,\cite{DTN85} and
thermo-\cite{HEN78} dissociation of water; dynamical properties of
the liquid,\cite{Fluc-q,POLARF} and anomalous
thermodynamics.\cite{MIS98}

There have been many previous attempts to develop potentials capable
of describing the various phases and properties of water, with
varying degrees of success. Comprehensive reviews are available in
Refs.~[\onlinecite{revu1,revu2}], and we will not attempt to review
the potentials in detail, but rather highlight essential features.
Most have concentrated on describing liquid water properties such as
the temperature-density variation, second virial coefficient,
diffusivity, radial distribution functions and structure functions;
others have focused on accurately reproducing gas-phase
spectrosocopic data.\cite{KES01} Some of the best-known potentials
are essentially empirical in nature (ST2,\cite{ST2} SPC,\cite{SPC}
SPC/E,\cite{SPC-E} TIP3P,\cite{TIP3P} TIP4P,\cite{TIP4P}
TIP5P\cite{TIP5P}), while others have used \textit{ab initio}
calculations carried out on small water clusters (monomer, dimer)
for their parameterizations (MCDHO,\cite{MCDHO} SAPT,\cite{SAPT}
NCC,\cite{NCC,NCC2} MCY,\cite{MCY} NEMO,\cite{NEMO}
CC-pol\cite{CC-POL}) or a combination of \textit{ab initio} and
experimental data (POL5,\cite{POL5} DIM water\cite{DIMWater}). In
almost all of the potentials, the parameterizations are carried out
with the implicit assumption that the basic structural unit consists
of the water monomer/molecule (notable exceptions being Halley {\it
et al.},\cite{Diss1} Corrales,\cite{Rene} and Voth {\it et
al.}\cite{EVB-Voth,MEVB})
Typically, the molecule is represented by a collection of point
charges placed at suitable sites so as to yield the correct dipole
and higher multipole moments for liquid water, as well as the
structures of small water clusters in some cases. The total energy
of a system comprised of water molecules is expressed as a sum of
coulombic and non-coulombic terms.
Simpler potentials hold the geometry of the water molecule as well
as the values of the point charges fixed\cite{SPC, TIP3P, TIP4P,
TIP5P, MB} while more realistic potentials allow OH bond
flexibility, modeled as harmonic and anharmonic
oscillators.\cite{ZRFP, FP} Further, rigid molecule models like that
of Dang and Chang,\cite{DC1} ASP,\cite{ASP} and NEMO\cite{NEMO}
include polarization effects by accounting for induced dipole
moments at every atom site in a self-consistent manner, while
potentials like TIP4P-FQ\cite{Fluc-q} use an approach similar to
ES+\cite{StrM} (see Section II.B) to account for polarization. Other
potentials that account for polarization effects include
MCDHO,\cite{MCDHO} which uses a three-site model in addition to a
negative mobile charge corresponding to a polarized electron cloud,
the diffuse charge pair potential model of Guillot and Guissani,
\cite{GG01} Polarflex,\cite{POLARF} based on empirical valence bond
theory, and TTM,\cite{TTM} which uses smeared charges and dipoles.

\subsection{The embedded-atom method and charge- dependent extensions}

The EAM formulation and extensions such as MEAM
have been used to successfully model a wide range of condensed phase
systems, including {\it fcc} metals,\cite{fcc} binary
alloys,\cite{alloys,PuGa} tin,\cite{RAB97-Sn} group IV elements such
as Si,\cite{MEAM} and even organic polymers.\cite{polyeth} In the
basic method, the total cohesive energy of a system is expressed as
a function of a local electron density, with each atom viewed as an
impurity embedded in a host consisting of the remaining atoms. The
host electron gas provides both ion-ion interactions and a
volume-dependent energy component.\cite{EAM} The total energy of the
system is written as follows:
\begin{equation}
E_{\rm EAM}=\sum_{i=1}^{N}{E_i}, \label{one}
\end{equation}
where $N$ is the number of atoms and
\begin{equation}
E_i=F_{i}(\overline{\rho}_i)+\frac{1}{2}\sum_{j\ne i}
\phi_{ij}(R_{ij}). \label{eam}
\end{equation}
$F_{i}$ is an element-dependent {\it embedding function} of the
effective local electron density $\overline{\rho}_i({\bf R}_i)$ at
atomic site $i$, and represents the collective many-body effects of
the remaining atoms in the host material; $\phi_{ij}$ corresponds to
a pair potential between interacting atoms $i$ and $j$. The
inclusion of the many-body term $F_i$ in the energy expression makes
the EAM significantly different from traditional two-body
potentials.  Moreover, in contrast to earlier approaches that
utilized bulk volume corrections, the EAM volume dependence is local
to each atom,\cite{EAM2} corresponding to an effective dependence on
local coordination.\cite{Payne93} Equivalently---in a tight-binding
bond picture of the EAM---the embedding energy can be understood in
terms of a local moment approximation to the density of
states.\cite{Voter94}

In the simplest EAM formulation, parameterized isolated-atom
electron densities are associated with each nucleus, and
$\overline{\rho}_i({\bf R}_i)$ is approximated by the sum of the
tails of all neighboring atom electron densities at site $i$:
\begin{equation}
\overline{\rho}_i({\bf R}_i) \simeq \sum_{\stackrel{j=1}{j\ne
i}}^{n_i}{\rho_j^a({\bf R}_{ij})}. \label{rhobar}
\end{equation}
Here $n_i$ is the number of nearest neighbors of atom $i$, and
$\rho_j^a$ corresponds to the isolated atomic electron density of
neighbor $j$. Various refinements of $\overline{\rho}_i$ have been
devised to account for neutral electron density
polarization\cite{MEAM,Daw89,WEG01} as well as the inclusion of
alloying effects.\cite{PuGa}

As noted in the Introduction, EAM-based potentials in their original
formulation do not account for explicit charge dependence or charge
transfer. To address this limitation in their models of metal-oxide
systems, Streitz and Mintmire\cite{StrM} proposed an extension of
EAM, ES+, in which an electrostatic energy term $E_{es}$ was added
to $E_{\rm EAM}$. $E_{es}$ is defined by the equation
\begin{equation}
E_{es}=\sum_{i=1}^{N}{{E_i}^{\rm ion}(q_i)} + \frac{1}{2}
\sum_{\stackrel{i,j=1}{i\ne j}}^N V_{ij}, \label{streitz}
\end{equation}
where ${E_i}^{\rm ion}(q_i)$ represents the ionization energy of an
isolated atom $i$, $q_i$ is its charge, and $V_{ij}$ is the coulomb
interaction energy. Following the Rapp\'e and Goddard QEq
model,\cite{RAG91} ${E_i}^{\rm ion}(q_i)$ is expressed in terms of
atomic charge $q_i$, atomic electronegativity $\chi_i^0$, and atomic
hardness $J_i^0$ via a second order Taylor series expansion about
the isolated neutral atom energy ${E_i}^{\rm iso}(0)$:
\begin{equation}
{E_i}^{\rm ion}(q_i)={E_i}^{\rm
iso}(0)+\chi_i^0q_i+\frac{1}{2}J_i^0{q_i^2}. \label{Eion}
\end{equation}
In addition, $V_{ij}$ is expressed in terms of effective electron
densities $\varrho_i$ and $\varrho_j$ of atoms $i,j$ as
\begin{equation}
V_{ij}=\int{\int{\frac{\varrho_i({\bf r},q_i;{\bf
R}_i)\varrho_j({\bf r}',q_j;{\bf R}_j)}{|{\bf r} - {\bf r}'|} \,
d{\bf r} \, d{\bf r}'}}, \label{Vij}
\end{equation}
where ${\bf R}_i$ and ${\bf R}_j$ are the position vectors of the
atomic nuclei, and $q_i$ and $q_j$ are the atomic charges. The
$\varrho_i$ include screened nuclear and polarized valence electron
components. The latter are modeled using a shape function $f_i$ with
a fixed (optimized) parametric form. The instantaneous charge on
each atom, which varies as a function of atomic configuration, is
obtained via chemical potential equalization, and requires the
solution of $N$ coupled linear equations involving the $V_{ij}$ and
a set of charge- and interaction-dependent electronegativities
$\chi_i$.

The original Streitz-Mintmire formulation was used to represent
atomic interactions in the Al-O system, and correctly predicted
elastic and energetic properties in the bulk, as well as surface energies
and relaxations, with reasonable assignments of ionic charges for the Al and O
atoms.\cite{StrM,StrM-surf} It was later used successfully in
dynamical simulations of the energetics of vacancies in $\gamma$-alumina\cite{STM04}
and in studies of the oxidation of aluminum nanoclusters,\cite{VAS} again with reasonable
values for the computed ionic charges. However, Zhou {\it al.} noted that
the model could not describe the behavior of the $\alpha$ phase of Al$_2$O$_3$ under
compression, wherein the computed charges oscillated between large unphysical
values at short interatomic spacings.\cite{CTIP}  This behavior was attributed to
a compensating effect on the part of the EAM component of the ES+ potential, whose
particular parameterization effectively constrained the atoms from approaching too closely.
To address this problem, and also enable the use of alternative EAM parameterizations
within ES+, Zhou {\it et al.} developed a variant in which {\it a priori}
empirical charge bounds were imposed on the ions in the
electrostatic component. The resulting CTIP-EAM model successfully
described cohesive and surface energies, surface oxidation, and thin-film growth
of various Al/Zr-oxide systems.

A primary limitation of both ES+ and CTIP-EAM is that they assume a
quadratic Taylor series expansion about the nominal ionic charges
and are thus valid only for reasonably small fluctuations about
these values.\cite{vat06,atv06} This precludes a non-perturbative
description of charge transfer in reactive systems, and the
significant electron density rearrangements that are induced by
strong intermolecular interactions. It also prevents a proper
description of the dissociation of interacting atomic and molecular
species, since the imposed quadratic dependence on charge does not
transition smoothly to the correct linear dependence at long
range.\cite{vat06a,PPLB,PerdewNATO}

A second pressing issue is the lack of self-consistency in both ES+
and CTIP-EAM, since the $\varrho_i({\bf r},q_i;{\bf R}_i)$ appearing
in the electrostatic component of these potentials is regarded as
formally distinct---and is parameterized separately from---the EAM
electron density $\rho_i^a$.
This must be regarded as problematic in light of the intrinsic
long-range, many-body nature of charge polarization and charge
transfer.  At the electronic level, it is well known that subtle
interactions in the vicinity of quantum mechanical curve
crossings,\cite{MultiState} and the concomitant interplay between
short and long-range electronic correlations, can have a profound
effect on the details of chemical bonding. Indeed, such effects in
water have been recently the focus of considerable theoretical and
experimental interest.\cite{H2ODissoc}  Both the problem of
significant charge polarization as well as the self-consistency
issue in ES+ have been noted previously in the context of
alumina.\cite{KNS98} At the atomistic level, these intrinsically
quantum mechanical effects must be properly reflected in the design
of the potential if it is to accurately describe charge transfer and
reactive dynamics.

\subsection{Charge-transfer embedded atom method potential for water}

A potential that addresses both of these issues within a density
functional-based multiscale formalism has been developed
recently.\cite{vat06,atv06} The formalism unifies all extant
embedded-atom models within a common theoretical framework, and as
an immediate consequence, generalizes to a fully-interacting,
self-consistent charge-transfer embedded-atom potential.  This
potential is used here as the starting point for constructing a new
charge-transfer potential for water.

For details, we refer the reader to the original papers. Here we
summarize the central results.  First, we note that the formalism
automatically imposes the requirement that $\rho$ equal $\varrho$ in
the embedding and electrostatic components of the potential, and
incorporates a proper treatment of the long-range dissociation of
interacting subsystems.\cite{vat04,vat06a}  These constraints
together effect the crucial balance between short- and long-range
electronic correlations.

The general CT-EAM form has been shown to be derivable from the
exact quantum-chemical atom-in-molecule (AIM)\cite{AIM} and
diatomics-in-molecule (DIM)\cite{DIM} Hamiltonians. In this picture,
charge-transfer-dependent embedding functions correspond to one-atom
AIM terms, while pair potentials map onto two-atom DIM
terms.\cite{vat06} This reformulation suggests a practical approach
to the parameterization of CT-EAM molecular potentials based on
resonance state (diabatic charge state) potential
curves.\cite{vat06b} Here, we adopt the parameterization perspective
of Refs.~[\onlinecite{atv06,vat04,vat06a}], which emphasizes the
charge-transfer electron densities as fundamental variables.
Regardless of which approach is chosen, however, two
key elements of the original theory must be modified: the form
assumed by the background embedding densities, and the total
cohesive energy expression.

Consider the background embedding density first. We begin by
decomposing the total electron density into a sum of AIM components,
denoted by $\rho_i^*({\bf r};{\bf R}_i)$.
Here ${\bf r}$ represents an arbitrary point in space for the
electronic coordinate, and the $\rho_i^*$'s are assumed centered on
the corresponding atomic nuclei.  The $\rho_i^*$ play an analogous
role in CT-EAM to the isolated electron densities $\rho_i^a$ in the
original EAM.  We use `atom-in-molecule' as a general term for
referring to the $\rho_i^*$, whether derived from molecules,
clusters, or solids. In principle, any physically-justified AIM
decomposition can be used.\cite{AIM,Bader} The sole requirement is
that the decomposition satisfy $\rho({\bf r}; {\bf R}) = \sum_i
\rho_i^*({\bf r}; {\bf R}_i)$, where $\bf R$ = $\{{\bf R}_j\}$ is a
collective variable representing the instantaneous geometry of all
atoms in the system.

The CT-EAM forms for the background embedding densities and
effective atomic charges $q_i$ are obtained as appropriate weighted
averages of the density difference
\begin{equation}
\Delta_{i}({\bf r}; {\bf R}) = \rho({\bf r}; {\bf R}) - \rho_i^a
({\bf r}; {\bf R}_i) \, . \label{drho}
\end{equation}
$\Delta_i$ corresponds to the electron density distribution of the
medium in which the $i$th atom is embedded. Let $\chi_{i}^{L}$ be a
weight function yielding the spatially-averaged quantity
$\Theta_{i}^{L}$:
\begin{equation}
   \Theta_{i}^{L} = \int \Delta_i({\bf r}; {\bf R})\, \chi_{i}^{L}({\bf r})\, d{\bf r}\, .
   \label{thetaL}
\end{equation}
If $\chi_{i}^{L}$ is constructed so as to project out $
\rho_{i}^{*}({\bf r};{\bf R}_i)$ from $ \rho({\bf r}; {\bf R})$, we
obtain a uniform average of the density difference between
$\rho_i^*$ and $\rho_i^a$, which is simply the effective charge:
\begin{equation}
q_i = \int \left( \rho_{i}^{*}({\bf r}; {\bf R}_i) - \rho_i^a({\bf
r}; {\bf R}_i) \right)\, d{\bf r}\, .
   \label{chargeanddensity}
\end{equation}
$q_{i}$ is the {\it localized} zeroth order moment $(L = 0)$ of
$\Delta_{i}({\bf r}; {\bf R})$. Note that this relation formally
connects the AIM densities and effective charges. This is the
mechanism through which CT-EAM imposes its requirement on the
embedding and electrostatic components of the potential, that
$\rho = \varrho$.

If instead we take $\chi_{i}^{L}$ equal to a $\delta$-function
centered on atom $i$, and utilize the density decomposition of
$\rho({\bf r})$, we obtain an expression for the embedding density
$\overline{\rho}_i^*$:
\begin{eqnarray}
\overline{\rho}_i^*({\bf R}_i) &=& \int \left[ \rho({\bf r}; {\bf
R})-\rho_i^a({\bf r};{\bf R}_i)\right] \delta({\bf r} - {\bf
R}_{i})\, d{\bf r} \nonumber  \\
&\approx& \int \left[ \rho({\bf r}; {\bf R})-\rho_i^*({\bf r};{\bf
R}_i)\right] \delta({\bf r} - {\bf R}_{i})\, d{\bf r} \nonumber  \\
&\approx& \sum_{j \ne i} \rho_{j}^*({\bf R}_{ij}). \label{rhobarij}
\end{eqnarray}
In the second step, we have approximated $\rho_i^a$ by $\rho_i^*$.
This is reasonable because for purposes of estimating the embedding
density, the difference between the isolated and AIM densities for
the atom experiencing the embedding is comparatively small. Most
contemporary EAM calculations already implement a similar
approximation: parameterized functional forms for the $\rho_j^a$'s
in Eq.~(\ref{rhobar}) are included within the overall potential
specification, and thus effectively serve as AIM $\rho_j^*$'s.

In light of Eq.~(\ref{rhobarij}), the CT-EAM background embedding
density $\overline{\rho}_i^*({\bf R}_i)$ at atom $i$ corresponds to
the localized infinite moment $(L = \infty)$ of $\Delta_{i}({\bf
r}; {\bf R})$.  $q_i$ and $\overline{\rho}_i^*({\bf R}_i)$ are thus
closely related, each expressible as a distinct localized moment of
$\Delta_{i}({\bf r}; {\bf R})$.

The second modification of the EAM concerns the cohesive energy
expression. The CT-EAM generalization of
Eqs.~(\ref{one})-(\ref{eam}) is\cite{vat06,atv06}
\begin{equation}
E = \sum_{i} \bigg[ \sum_{M=1}^{M_i} \Omega_{i,M} F_{i,M}
[\overline{\rho}_{i,M}^*] + \frac{1}{2}\sum_{j \ne i}
\sum_{M=1}^{M_{ij}} \Omega_{ij,M} \Phi_{ij,M}\bigg].
\label{finalexpression}
\end{equation}

\vspace*{.1in}

\noindent In the embedding term, the index $M$ sums over the $M_i$
integer charge states that are to be included in the model for the
$i$th atom; in the pair interaction term, $M$ sums over all $M_{ij}$
{\it pairs} of included charge states.  $F_{i,M}$, $\Phi_{ij,M}$ and
$\overline{\rho}_{i,M}^*$ are charge-transfer generalizations of the
conventional EAM quantities. The $\Omega_{i,M}$ and $\Omega_{ij,M}$
are weighting factors for the particular integer charge states or
combinations of charge states that are instantaneously populated for
a given system configuration.

In order to make practical use of the CT-EAM formulation, it is
necessary to choose the number of charge states to be included for
each atom type, and also the parametric functional forms to be used
for the embedding functions $F_{i,M}$ and pair interactions
$\Phi_{ij,M}$. These choices are discussed below in Section III.

\section{CT-EAM potential for water}

An obvious concern with the fixed-charge models is that they lack
the flexibility to describe phenomena where the neutral water
molecule is not necessarily the fundamental structural unit. Even
more sophisticated approaches such as MEVB and SAPT associate
charges with fixed molecular and ionic species (water, hydronium
ion). Additionally, they rely on the specification of appropriate
quantum-mechanical basis states in order to compute dynamical
charges. These features make such models difficult to generalize to
the study of larger and more complex water-containing systems where
charge transfer effects are expected to play a significant role.
Important examples include the dynamics of solvated
proteins,\cite{PDyn} water-silica interactions,\cite{H2O-silica}
energy transduction in molecular motor proteins,\cite{Cui06} and the
electronic and magnetic properties of exotic materials.\cite{NMAT}

The use of the EAM as the starting point of our approach means that
our perspective is shifted from larger molecular building blocks to
a more fine-grained picture---exact in DIM---of perturbed atoms
embedded in a many-body medium, and explicit two-body interactions. The
formal basis for the methodology in density functional theory
implies that CT-EAM potentials are in principle capable of
describing arbitrary charge states and energetics of the atoms in
any given local chemical environment.

Using Eq.~(\ref{finalexpression}) as our starting point, we will now
develop an environment-dependent potential that is parameterized to
reproduce the ground-state energy and geometry of the water monomer
and dimer for select geometries of these structures. The
parameterization incorporates charge transfer information derived
from {\it ab initio} calculations on the hydronium and hydroxyl
ions, the neutral isolated water molecule, and neutral water dimer.

\subsection{Environment-dependent atomic charges}
In principle, the CT-EAM potential should be formulated in terms of
AIM electron densities and a relatively complete set of atomic
charge states, as outlined in the previous section. In this first
application of the theory, however, our aim is to explore the
capabilities of the framework in the simplest possible
implementation.  We therefore adopt the AIM atomic charge as a
surrogate for the background density at a given atomic site;
ultimately, it will be necessary to utilize more detailed
approximations of the AIM spatial distributions, particularly for
dynamical simulations. As is clear from the discussion surrounding
Eqs.~(\ref{chargeanddensity})--(\ref{rhobarij}), this
approximation is equivalent to replacing the electron density
distribution by its localized zeroth order moment. We also assume
two charge states per atom, as discussed below.

Given the functional form for the CT-EAM energy
(Eq.~(\ref{finalexpression})), our first task is to develop
appropriate parameterizations for the atomic charges $q_i$.  The
dataset used to fit the AIM charges is computed using standard
population analysis techniques in conjunction with {\it ab initio}
calculations.  We have used the {\it ab initio} software package
{\tt GAMESS}\cite{gamess} at the unrestricted Hartree-Fock (UHF)
level and with a fairly high-quality basis set---6-31G**---to obtain
the L\"{o}wdin atomic charges.\cite{fn1}  The same basis set and
level of theory were used in all parameterization calculations
throughout this work, and all charge and potential parameters were
varied in order to limit model estimation errors to less than 0.015
$e$ and 0.02 eV per molecule, respectively. Although electron
correlation and other effects such as zero-point energy corrections
are not included in these calculations, Maheshwary {\it et al.} have
performed extensive {\it ab initio} calculations on water clusters
using HF/6-31G** and concluded that overall trends in the variation
of energy with cluster size remained unaltered with further
improvements in basis set and level of theory.\cite{ab2} Indeed,
comparison of their results with recent accurate
X3LYP hybrid density functional energies, computed with a much larger
aug-cc-pVTZ(-f) basis set,\cite{Su04} reveals a nearly identical pattern of variation
in stabilization energy for the most stable geometry of ${\rm (H}_2{\rm O)}_n$
as a function of cluster size $n$. As our model systems we choose different geometries of:
i) the neutral water molecule, ii) the hydronium ion ${\rm H}_3{\rm O}^+$, iii) the
${\rm OH}^-$ ion, and iv) the water dimer, in order to represent diverse
coordination environments.  It is
important to bear in mind that these structures and geometries are
used here to parameterize charge rather than energy. We therefore
expect electron correlation effects to be less important than the
quality of the basis set.

The motivation behind using different model systems is to ensure
that the resulting interatomic potential is sufficiently robust to
describe different chemical environments that the oxygen and
hydrogen atomic species might encounter in various water polymorphs.
The particular choice of the ${\rm H}_3{\rm O}^+$ and ${\rm OH}^-$ ions
is based on two key considerations. First,
they provide a coordination environment for the hydrogen and oxygen
atoms that is distinct from neutral interacting ${\rm H}_2{\rm O}$
dimers.  Second, ${\rm H}_3{\rm O}^+$ and ${\rm OH}^-$ are the two
primary dissociation products of water in solution, and thus are
essential to describing chemical reactions involving water.

A distinguishing feature of our methodology is the identification of
\emph{local clusters} within an instantaneous configuration of the
system, and the indexing of an atom's charge based on the kind of
cluster to which it belongs. Each local cluster is assigned a charge
depending on the number of atoms in the cluster, with the charge
partitioned among the cluster atoms in a geometry-dependent manner.
For each oxygen atom, we identify the number of hydrogen atoms
within a radius that is chosen to be 1.5 \AA.  For example, if two
hydrogen atoms are in close proximity to an oxygen atom, then the
cluster (oxygen plus two hydrogen atoms) constitutes a neutral ${\rm
H}_2{\rm O}$ cluster, while if the number of hydrogens surrounding
an oxygen atom is three, then the cluster is identified as a ${\rm
H}_3{\rm O}^+$ with a net cluster charge of $+1$. The total charge on
an identified cluster with
$N_{\rm H}$ hydrogen atoms is thus $N_{\rm H}-2$. Once all clusters have
been identified, the total cluster charge is partitioned among the
atoms as a function of their relative positions within the cluster.
We then account for further charge polarization and charge transfer
between neighboring clusters by parameterizing the amount of charge
transferred between two water monomers (constituting a dimer) as a
function of the hydrogen-bond distance between the two monomers. The
final total charge on a given atom consists of both intra-cluster
and inter-cluster contributions; this corresponds to its effective
AIM charge. The following section provides further details of the
charge parameterization procedure.

\subsection{Model Clusters: ${\rm H}_2{\rm O}$, ${\rm H}_3{\rm O}^+$,
and ${\rm OH}^-$}

\begin{figure}
\includegraphics[scale=.62]{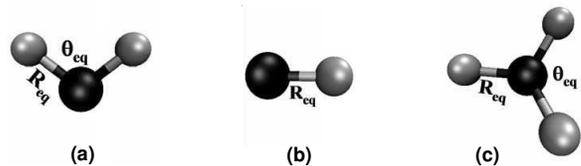}
\caption{Geometry of the three model clusters: (a) ${\rm H}_2{\rm
O}$; (b) ${\rm OH}^-$; (c) ${\rm H}_3{\rm O}^+$.} \label{fig:geom}
\end{figure}

For the three model clusters, we initially obtained the equilibrium
geometries as given in Table~\ref{tab:EquiTable}. Next, we varied
the geometries of the three systems to obtain the L\"{o}wdin atomic
charges as a function of system geometry as shown in
Fig.~\ref{fig:geom}. For the water molecule, atomic charges for the
three vibrational modes (symmetric stretch, asymmetric stretch, and
bending) were obtained, and the charges were fitted as a function of
the two OH bond distances and the intramolecular angle. The
symmetric and asymmetric stretches and contractions varied from 70\%
to 140\% of the equilibrium bond length ($R_{eq}$) at various values
(55\%--130\%) of the equilibrium intramolecular angle
($\theta_{eq}$). In a similar fashion, we obtained charges for
symmetric deformations (70\% to 140\%) of the OH bonds with the
three bond angles fixed at the equilibrium value for the ${\rm
H}_3{\rm O}^+$ ion as well as the L\"{o}wdin charges on the O and H
atoms for the ${\rm OH}^-$ anion for deformations ranging from 70\%
to 140\% of the equilibrium OH bond distance. We then fitted the
atomic charge variations as a function of the relative positions of
the respective atoms in the cluster.

\begin{table}
\caption{\label{tab:EquiTable} Equilibrium geometry values for the
model clusters.  Angles in degrees; distances in \AA.}
\begin{ruledtabular}
\begin{tabular}{lcc}
                     & $\theta_{eq}$          & $R_{eq}$          \\ \hline
${\rm OH}^-$         & ---                    & $0.958$           \\
${\rm H}_2{\rm O}$   & $105.5$                & $0.945$           \\
${\rm H}_3{\rm O}^+$ & $114.74$               & $0.961$           \\
\end{tabular}
\end{ruledtabular}
\end{table}

We now present the parameterization equations relating the variation
in atomic charge with respect to cluster geometry. As noted
previously, the number of atoms in a cluster is defined by a central
oxygen and the number of hydrogen atoms that lie within a specified
radial cutoff $r_{cut}$ = 1.5 \AA.

Consider a cluster with a central oxygen O and $N_{\rm H}> 1$
hydrogen atoms. Let the position vector of the $p^{th}$ hydrogen
atom with respect to the central oxygen atom O be ${\bf r}_p$. The
charge $q^p_{\rm H}$ on the $p^{th}$ hydrogen atom is expressed as a
function of the positions of all atoms in the cluster, specifically,
$\theta_{p{\rm O}s}$, $r_p$ and $r_s$, where $s$ corresponds to any
of the other $N_{\rm H}-1$ hydrogen atoms in the cluster,
$\theta_{p{\rm O}s}$ is the angle between ${\bf r}_p$ and ${\bf
r}_s$, and $r_s$ is the distance of the $s^{th}$ hydrogen atom from
O.  We have
\begin{equation}
 {q^p_H}=q^p_1+q^p_2+q^p_{N_{\rm H} > 2}, \label{totalHq}
\end{equation}
where
\begin{eqnarray}
q^p_1 &=& \sum_{\stackrel{s=1}{s\neq p}}^{N}
\Big[\alpha(\theta_{s{\rm O}p})e^{-2r_p} +
{\beta(\theta_{p{\rm O}s})} r_pe^{-r_p} \nonumber \\
&& \hspace*{.5in} +\ c(\theta_{p{\rm O}s})\Big]
\sin^{2}{\theta_{p{\rm O}s}}, \label{qp1}
\end{eqnarray}
\begin{equation}
q^p_2=\sum_{\stackrel{s=1}{s\neq p}}^{N_{\rm H}}{(r_p-r_s)
d(\theta_{p{\rm O}s})\sin^2{\theta_{p{\rm O}s}}}, \label{qp2}
\end{equation}
and
\begin{equation}
q^p_{N_{\rm H} > 2}=\sum_{\stackrel{s=1}{s\neq p}}^{N_{\rm
H}}\left[{\alpha}_2e^{-2r_p} +
{\beta}_2r_pe^{-r_p}+\gamma_2e^{-r_p}\right] \sin^{2}{\theta_{p{\rm
O}s}}. \label{qpn2}
\end{equation}
$q^p_{N_{\rm H} > 2}$ is non-zero when $N_{\rm H}>2$. $\alpha$
$\beta$, $c$, $d$ are functions of $\theta_{p{\rm O}s}$, defined in
Eqs.~(\ref{alpha})--(\ref{dee}) below; ${\alpha}_2$, ${\beta}_2$,
and $\gamma_2$ are constants whose values are given in
Table~\ref{tab:paramsI}. For the special case where the identified
cluster contains only a single hydrogen, the charge on the hydrogen
is given by
\begin{equation}
q_{\rm H}^p={\alpha}_1e^{-2r_p}+{\beta}_1e^{-r_p}+\gamma_1,
\label{qn1}
\end{equation}
where ${\alpha}_1$, ${\beta}_1$, and $\gamma_1$ are constants specified
in Table~\ref{tab:paramsI}.

To prevent energy discontinuities, we utilize a switching function
${\cal S}(t)$ to modulate the calculated charge on hydrogen atom $p$
as a function of $r_p$:
\begin{equation}
\tilde{q}_{\rm H}^p = q_{\rm H}^p\ {\cal S}(r_p-t_{cut}),
\label{qpmod}
\end{equation}
where
\begin{equation}
{\cal S}(t) \equiv \frac{1}{2}\Big(1-\tanh (t/t_1)\Big).
\label{switch}
\end{equation}
Here $q_{\rm H}^p$ is given by Eq.~(\ref{totalHq}) or
Eq.~(\ref{qn1}) depending on the oxygen coordination, $t_{cut}$ =
1.41 \AA , and $t_1$ is given in Table~\ref{tab:paramsI}. The role
of ${\cal S}$ is to asymptotically switch ${\tilde q}_{\rm H}^p$
from $q_{\rm H}^p$ to zero at a radius that is less than the cluster
assignment cutoff $r_{cut}$ (note that $t_{cut} < r_{cut}$). This
prevents energy discontinuities when an H crosses from outside to
inside the $r_{cut}$ boundary. Depending on the number of hydrogen
atoms $N_{\rm H}$ in the cluster, the charge $q_{\rm O}$ on the oxygen atom
in the cluster is
\begin{equation}
q_{\rm O}=(N_{\rm H}-2)-\sum_{p=1}^{N_{\rm H}}{\tilde q}_{\rm H}^p.
\label{qo}
\end{equation}

\begin{table}
\caption{\label{tab:paramsI} Water monomer charge model parameters I.}
\begin{ruledtabular}
\begin{tabular}{ccccccc}
$\alpha_1$ ($e$)  & $\alpha_2$ ($e$) & $\beta_1$ ($e$) & $\beta_2$ ($e$/\AA)&
$\gamma_1$ ($e$)  & $\gamma_2$ ($e$) & $t_1$ (\AA)      \\ \hline
$-10.032$         & $9.3087$         & $7.7183$        & $3.0558$           &
$-1.4124$         & $-6.7189$        & $0.03$ \\
\end{tabular}
\end{ruledtabular}
\end{table}

The functional forms for $\alpha$, $\beta$, $c$ and $d$ are given by
the following equations (parameters are listed in
Table~\ref{tab:paramsII}):
\begin{equation}
\alpha(\theta)=a_1e^{\theta}+a_2e^{\theta^2/4}+a_3\theta,
\label{alpha}
\end{equation}
\begin{equation}
\beta(\theta)=b_1\theta e^{-\theta}+b_2\theta e^{-2\theta} +
b_3{\theta^2}, \label{beta}
\end{equation}
\begin{equation}
c(\theta)=c_1\theta e^{-\theta}+c_2\theta e^{-2\theta}+c_3,
\label{cee}
\end{equation}
\begin{equation}
d(\theta)=d_1(\theta -{\theta}^2)+d_2{\theta}^3.  \label{dee}
\end{equation}

\begin{table*}
\caption{\label{tab:paramsII} Water monomer charge model parameters II.}
\begin{ruledtabular}
\begin{tabular}{ccccccccccc}
$a_1$ ($e$)             & $a_2$ ($e$)             & $a_3$ ($e$) &
$b_1$ ($e$)             & $b_2$ ($e$/\AA)         & $b_3$ ($e/{\rm
\AA}^2$)       & $c_1$ ($e/{\rm \AA }^2$) & $c_2$ ($e/{\rm \AA }^3$)
& $c_3$ ($e/{\rm \AA}^3$) & $d_1$ ($e$/\AA)         & $d_2$
($e$/\AA)
\\ \hline $0.8158$                & $-3.2198$ & $0.5725$
& $-14.0058$              & $61.8232$ & $1.3188$                &
$12.2353$               & $-15.7797$ & $-3.5928$               &
$0.4430$                & $0.1154$   \\
\end{tabular}
\end{ruledtabular}
\end{table*}

\begin{figure}
\includegraphics[scale=.32,angle=-90]{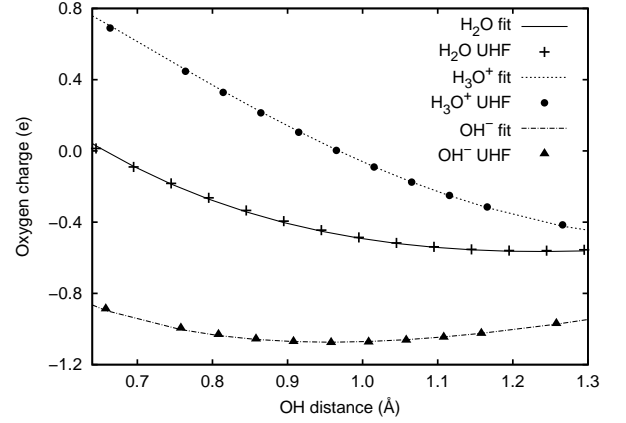}
\caption{\label{fig:OCharge} Oxygen charge as a function of OH
distance for a symmetric variation at $\theta_{eq}$ in ${\rm
H}_2{\rm O}$, ${\rm H}_3{\rm O}^+$, and ${\rm OH}^-$.}
\end{figure}

Fig.~\ref{fig:OCharge} compares the actual and predicted charges for
the oxygen atom in the three different clusters for select
geometries.  The fits are very good in each case.

\subsection{Charge Transfer between Clusters: Water Dimer
Atomic Charges }

\begin{figure}
\includegraphics[scale=0.17]{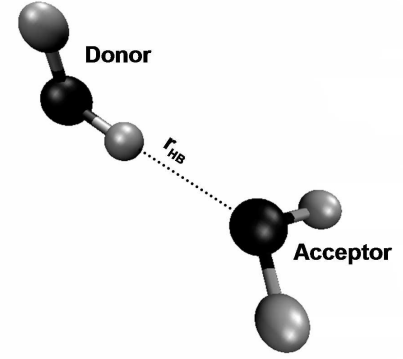}
\caption{\label{fig:dimer} Water dimer geometry, illustrating
hydrogen bonding.}
\end{figure}

Water polymorphs are characterized by the formation of hydrogen
bonds between neighboring water molecules. Thus, the environment of
any atom in bulk is different than when it is part of an isolated
water molecule. In order to include the effect of a bulk environment
on the atomic charges, we considered two water molecules (dimer) and
parameterized the atomic charges for select geometries of the dimer
(see Fig.~\ref{fig:dimer}). This was done by fixing the geometries
of the individual water molecules to match the equilibrium isolated
water geometry and moving the two molecules relative to each other
along the line of hydrogen bonding ${\bf r}_{\rm HB}$ between
the two molecules.

Using the same level of theory as above, the equilibrium dimer
geometry was computed, yielding intramolecular bond distance and
bond angles of 0.950 \AA\ and $105.5^{\circ}$, respectively. The
computed intermolecular hydrogen bond distance $r_{\rm HB}$ was
$2.03$ \AA, with an intermolecular bond angle $\varphi$ (formed between
the two oxygens and the common hydrogen; see Fig.~\ref{fig:dimfig})
of $172.3^{\circ}$. Note that the equilibrium intramolecular bond
distances and angles for each molecule are very similar to those of
a single water molecule, while the intermolecular angle corresponds
to a nearly linear configuration.

We next varied $r_{\rm HB}$ from 75\% to 130\% of its equilibrium
value, keeping the geometry of the two molecules rigid and fixing
$\varphi$ at its equilibrium value. This resulted in a finite
intermolecular charge transfer between the two molecules, such that
the donor molecule became negatively charged relative to the
acceptor molecule as a function of the common hydrogen position.
Based on these results, we defined a net intermolecular charge
transfer $dq$ (donor $\rightarrow$ acceptor) between the two
clusters, and a partitioning---referred to collectively as
\{$dq_i$\}---of this charge transfer among the constituent atoms.
$dq$ is parameterized as:
\begin{equation}
dq=a_{im}e^{-b_{im}r_{\rm HB}}, \label{IM}
\end{equation}
where $r_{\rm HB}$ is the distance between the donor hydrogen and
acceptor oxygen. Based on our HF calculations, we choose the cutoff
for charge transfer between clusters to be $s_{cut} = 2.5$ \AA\
(beyond this distance the computed $dq$ was effectively zero.) As
above ({\it cf.} Eq.~(\ref{qpmod})), the switching function ${\cal
S}$ modulates $dq$ so as to ensure energy continuity; it guarantees
that if $r_{\rm HB} > s_{cut}$, there is no intermolecular charge
transfer:
\begin{equation}
d{\tilde q} = dq\ {\cal S}(r_{\rm HB}-r_{qim}). \label{dqIM}
\end{equation}
The parameters $a_{im}$, $b_{im}$, and $r_{qim}$ in Eqs.~(\ref{IM})
and (\ref{dqIM}) are given in Table~\ref{tab:dparams}. $d{\tilde q}$
is partitioned among the atoms as follows:
\begin{equation}
dq^{{\rm O}_{donor}}=-0.5\,d{\tilde q} \label{Odonor}
\end{equation}
\begin{equation}
dq^{{\rm O}_{acceptor}}=0.75\,d{\tilde q}, \label{Oaccpt}
\end{equation}
and
\begin{equation}
dq^{{\rm H}_{donor}}= -0.4\,d{\tilde q},  \label{Hdonor}
\end{equation}
where ${\rm O}_{donor}$, ${\rm O}_{acceptor}$, and ${\rm H}_{donor}$
represent the donor oxygen, acceptor oxygen, and donor hydrogen
respectively. For the acceptor molecule, $dq^{{\rm H}_{acceptor}}$
is computed by partitioning $[d{\tilde q} - dq^{{\rm
O}_{acceptor}}]$ equally among the constituent hydrogens in that
cluster.  Similarly, $[(dq^{{\rm O}_{donor}} + dq^{{\rm H}_{donor}})
- d{\tilde q}]$ is distributed equally among the hydrogen atoms
(other than ${\rm H}_{donor}$) in the donor cluster. The total
charge $q_i$ on the $i$th atom is given by
\begin{equation}
q_i = q_i^{cl} + dq_i, \label{TOTALQ}
\end{equation}
where $q_i^{cl}$ is the atomic charge due to intramolecular charge
transfer, calculated from Eqs.~(\ref{totalHq})--(\ref{qo})
($q_i^{cl} = q_{\rm H}^p$ for H, and $q_{\rm O}$ for O), and $dq_i$
is the additional atomic charge acquired via intermolecular charge
transfer (Eqs.~(\ref{IM})--(\ref{Hdonor})). As we shall see shortly,
all three values---$q_i$, $q_i^{cl}$, and $dq_i$---are needed for
computing the total energy in CT-EAM.  $q_i^{cl}$ and $dq_i$ are the
background density arguments to distinct charge transfer embedding
functions (see Section III.D), and $q_i$ is used to compute the
coulomb pair interaction energy.

\begin{table}
\caption{\label{tab:dparams} Water dimer charge model parameters.}
\begin{ruledtabular}
\begin{tabular}{lcr}
$a_{im}$ ($e$)  & $b_{im} ({\rm \AA}^{-1})$  & $r_{qim}$ (\AA) \\ \hline
$1.5812$        & $1.8222$                   & $2.35$          \\
\end{tabular}
\end{ruledtabular}
\end{table}

The specification of the CT-EAM charge transfer parameterizations
for both intra- and inter-cluster interactions is now complete. Note
that we have assumed that each cluster is defined such that a given H
atom belongs to only one cluster.  In particular, the identification
of a `hydrogen-bonding' H atom implicitly assumes that the H
atom belongs to one cluster and is hydrogen-bonded to the oxygen of
the neighboring cluster.

There are a number of special cases that may arise; these are
handled as follows. If a hydrogen belongs to more than one cluster,
we initially treat the clusters separately, account for their
cluster charges, and add the respective contributions for the common
hydrogen. Then we use the inter-cluster charge transfer function to
determine the charge transfer between the two clusters in each
direction, and add the results. That is, for the given pair of
clusters, we consider both scenarios where one cluster acts as a
donor and the other as an acceptor and vice-versa. If more than one
donor hydrogen is shared between two clusters, we use the same set
of charge transfer equations to compute two sets of charge transfers
between the clusters: there is no coupling between them. Finally,
the charge transfer between clusters is always mediated by the
\emph{hydrogen} atoms, irrespective of the relative distances
between the corresponding oxygens.

\subsection{Charge-dependent embedding functions}
We have described two intrinsic types of charge transfer in the
water system---inter- and intra-molecular---and presented
parameterizations for each.  These two types of charge transfer make
distinct contributions to the energy through their respective
charge-state-dependent embedding functions ({\it cf.}
Eq.~(\ref{finalexpression})). We must now consider how to determine
appropriate functional forms for these embedding functions.

In the original EAM formulation, the atomic embedding functions were
determined by numerical fits of the energy to configurational
reference states along a symmetric dilatation curve.\cite{EAM-rev}
Later, as a key aspect of MEAM, Baskes proposed the use of a
universal $\rho \ln \rho$ functional form, with the density argument
normalized to a reference state. Baskes rationalized this form by
noting that it gave the correct coordination dependence between bond
length and energy (bond-order/bond-length correlation) for
Si.\cite{MEAM} Indeed, MEAM has since proved remarkably robust in
applications to chemically-diverse materials
systems.\cite{EAM-rev,MEAM,alloys,RAB97-Sn} This suggests that the
same form may also work well as an {\it ansatz} for the
charge-transfer embedding functions required here.

An independent rationale for the $\rho \ln \rho$ form comes from
recent work on ensemble models of charge transfer for
strongly-interacting subsystems.\cite{vat06a}
In a resonance-state (microscopic) ensemble picture, the equilibrium
charge transfer within a larger closed system provides a measure of
the interaction strength between subsystems. In the equivalent
thermodynamic ensemble, the charge transfer parameter maps onto a
{\it non-zero} electronic temperature. This temperature is conjugate
to the charge-density entropy induced by the electronic polarization
and charge transfer among constituent subsystems. Interpreting the
charge transfer in terms of an effective electronic temperature
suggests using the information-theoretic form of the entropy
($\sum_i \rho_i \ln \rho_i$, where $i$ indexes the pure states
contributing to the ensemble\cite{PerdewNATO}), to model the
charge-transfer embedding energies.

In light of the universal nature of the density functional
electronic theory underlying CT-EAM, we expect the embedding
functional form to be independent of the nature (inter- or
intra-molecular) of the charge transfer. We therefore adopt the
$\rho \ln \rho$ form for all charge-transfer embedding functions---using
charges instead of background densities as discussed below.
Finally, the ensemble formulation and information-theoretic
interpretation both suggest that distinct charge transfer
contributions should enter additively into the overall energy
expression; this is consistent with the formal result in
Eq.~(\ref{finalexpression}).

\subsection{Embedding function and pair interaction parameterizations}

The parameterizations we have chosen for the AIM charges imply a
choice of $M_i = 2$ for both H and O, and $M_{ij} = 3$ for the pair
interactions (Eqs.~(\ref{phioo})--(\ref{phihh})). We write the net
embedding energy contribution $F_i$ of the $i$th atom
in terms of $q_i^{cl}$ and $dq_i$ as:
\begin{equation}
F_i=A_{i}q_i^{cl}\ln(e_0{q_i^{cl}}^2)+A^d_{i}dq_i
\ln(e_0{dq_i}^2),
\label{Fofrhobar}
\end{equation}
where $e_0 = 1$ has dimensions of $e^{-2}$. We use the square of the
charge to ensure a positive argument for the logarithm; the
additional factors of two are absorbed into the parameterization via
the prefactors.  The intra- and interatomic charge transfer values
$q_i^{cl}$ and $dq_i$ are used in lieu of the nominal background
embedding densities $\overline{\rho}_{i,M}^*$, $M$ = 1,2.
The justification for this comes from the common origin of $q_i$ and
$\overline{\rho}_i^*$ in Eq.~(\ref{thetaL}). We have also absorbed the
weighting factors $\Omega_{i,M}$ and $\Omega_{ij,M}$ into our
parameterizations (Eqs.~(\ref{Fofrhobar})--(\ref{phihh})).

Further insight into Eq.~(\ref{Fofrhobar}) can be obtained by
regarding the first term as corresponding to conventional EAM, with
the AIM charges within the water monomer playing the role of the
embedding electron density. This term is associated with
first-neighbor, intra-molecular charge transfer. The second term is
then a CT-EAM correction for second nearest-neighbor,
inter-molecular charge transfer. It is interesting to note in this
connection that a second-nearest-neighbor MEAM has been proposed
recently, aimed at correcting the structural stability and surface
energy orderings in certain {\it bcc} metals.\cite{MEAM-2NN}

The total pair interaction $\Phi_{ij}$ is given by the sum of two
terms, a classical electrostatic component, $V_{ij}$ (analogous to
Streitz and Mintmire's $V_{ij}$, {\it cf.} Eq.~(\ref{streitz})), and
a {\it non-coulombic} component, $\phi_{ij}$ ({\it cf.}
Eq.~(\ref{eam})):
\begin{equation}
\Phi_{ij}=  V_{ij} + \phi_{ij}. \label{PhiEq}
\end{equation}
Since we have chosen to utilize localized zeroth-order moment models
of the AIM electron densities, the electrostatic component of
$\Phi_{ij}$ consists simply of the classical coulombic interaction
between AIM charges $q_i$ and $q_j$,
\begin{equation}
V_{ij}= q_iq_j/R_{ij}. \label{VijEq}
\end{equation}
These charges are constrained to be identical to those appearing in
the embedding component of the potential, in accordance with the
CT-EAM self-consistency requirement. The form of the non-coulombic
potential is dictated by energy fits once the charge-dependent
components have been determined. These assume a purely repulsive
Born-Mayer-type form for the homonuclear pair interactions, and a
linear-exponential form for the OH interaction.   They are similar
to the functional forms utilized for pair interactions in the
original EAM,\cite{fcc} and are given by:
\begin{equation}
\phi_{\rm OO}=a_{\rm OO} e^{-4r_0r_{\rm OO}}, \label{phioo}
\end{equation}
\begin{equation}
\phi_{\rm OH} = 2 \left[a_{\rm OH} r_{\rm OH} + b_{\rm OH}
e^{-r_0r_{\rm OH}} + \frac{c_{\rm OH}}{r_{\rm OH}^{24}}\right] {\cal
S}(r_{\rm OH}-r_{cut}), \label{phioh}
\end{equation}
and \vspace*{-.15in}
\begin{equation}
\phi_{\rm HH} = 2a_{\rm HH} e^{-2r_0r_{\rm HH}} {\cal S}(r_{\rm
HH}-H_{cut}). \label{phihh}
\end{equation}

\vspace*{.01in}

\noindent

\begin{figure}
\vspace*{.25in}
\includegraphics[scale=.48,angle=0]{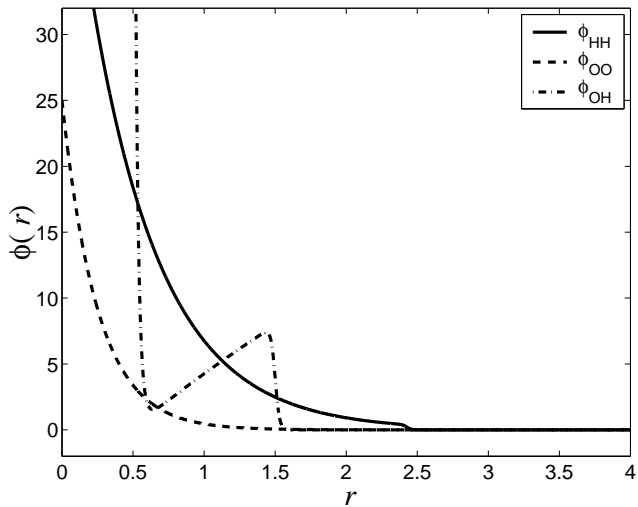}
\caption{\label{fig:Phi} Pair interaction potentials (in eV) as a
function of internuclear separation $r$ (in \AA ), for H-H, O-O, and
O-H.}
\end{figure}

In these expressions, $r_0 =1$ has dimensions of ${\rm \AA}^{-1}$,
and ${\cal S}(t)$ is the switching function defined in
Eq.~(\ref{switch}). $\phi_{\rm HH}$ and $\phi_{\rm OO}$ are purely
repulsive. They damp to zero beyond their respective cutoffs
$r_{cut}$ and $H_{cut}$ (the latter is specified in
Table~\ref{tab:Eparams}).  For consistency, $r_{cut}$ is taken to be
the same value as used above for determining whether an H atom belongs
to a particular cluster. This prevents non-coulombic H-H and O-H
interactions between atoms in different clusters, for geometries
near equilibrium. Note that $\phi_{\rm OH}$ has been designed to be
very repulsive at small O-H separations by including an $r_{\rm
OH}^{-24}$ term; this prevents the appearance of unphysical energy
minima.  The pair potentials are plotted in Fig.~\ref{fig:Phi}. The
unusual ``coat-hanger" shape of $\phi_{\rm OH}$ is a consequence of
the fact that the pair potentials are parameterized in conjunction
with the electrostatic term $\phi_{ij}$, as part of an overall fit
({\it cf.} Eqs.~(\ref{PhiEq}) and (\ref{VijEq})).  The particular
shape prevents O--H interactions between neighboring clusters.

The parameterization of the embedding functions and non-coulombic
interactions was carried out with respect to a set of reference
energies. We chose the symmetric mode of the monomer for three
different bond angles ($105.9$, $100$, and $110^\circ$) and the
equilibrium geometry of the dimer as our reference configurations.
Energies at each geometric configuration were obtained by
subtracting the isolated atom energies from the total energy
obtained via \textit{ab initio} UHF 6-31G** calculations using {\tt
GAMESS}. The energy of the isolated oxygen computed using this basis set was
$-74.7839$ hartrees and that of the isolated hydrogen atom equaled
$-0.5$ hartrees. $A^d_{\rm O}$ and $A^d_{\rm H}$ are constants given
in Table~\ref{tab:Eparams}.

For an oxygen atom O in a cluster with $N$ hydrogens ($N\geq$ 2),
\begin{eqnarray}
A_{\rm O} &=& -A_{\rm
EO}\sum_{j=1}^N\sum_{k=j+1}^N\sin^{2}(\theta_{j{\rm O}k})
\nonumber \\
&& \times\  \exp \big[ -\small{\frac{1}{2}} r_0^2
\left(r_{{\rm O}j}-r_{{\rm O}k}\right)^2 \sin^{2}(\theta_{j{\rm O}k})\big], \\
\nonumber \label{AO}
\end{eqnarray}
where $j$ and $k$ represent the $j^{th}$ and $k^{th}$ hydrogens in
the cluster, and $A_{\rm EO}$ is defined in Table~\ref{tab:Eparams}.
If $N=1$, we set $A_O = 2A_{\rm EO}$.
\begin{table*}
\caption{\label{tab:Eparams} Energy parameters. $a_{\rm OO}$,
$a_{\rm OH}$, $a_{\rm HH}$, and $b_{\rm OH}$ in eV; $c_{\rm OH}$ in
units ${\rm eV} \cdot {\rm \AA}^{24}$; $A_{\rm EO}$, $A^d_{\rm O}$,
$A_{\rm EH}$, and $A^d_{\rm H}$ in eV/$e$; $H_{cut}$, $r_{hs}$, and
$t_2$ in \AA; $\eta$ is dimensionless. }
\begin{ruledtabular}
\begin{tabular}{ccccccccccccc}
$a_{\rm OO}$ & $a_{\rm OH}$  & $a_{\rm HH}$  & $b_{\rm OH}$ &
$c_{\rm OH}$ & $A_{\rm EO}$  & $A^d_{\rm O}$ & $A_{\rm EH}$ &
$A^d_{\rm H}$ & $H_{cut}$  & $r_{hs}$ & $t_2$ & $\eta$ \\ \hline
$25.0$ & $3.0111$ & $25.0$ & $-2.4053$ & $2.5 \times 10^{-6}$ &
$-11.429$ & $0.0$  & $4.7621$ & $-0.5$ & $2.43$    & $2.1$         &
$0.1$ &
$0.0505$ \\
\end{tabular}
\end{ruledtabular}
\end{table*}

If a hydrogen atom $p$ that belongs to a cluster containing the
oxygen atom $s$ is involved in hydrogen bonding with $N_{\rm OH}$
oxygens of $N_{\rm OH}$ different neighboring clusters, then
\begin{widetext}
\begin{equation}
A_{\rm H}=A_{\rm EH}\Bigg[1 + \eta \sum_{u=1}^{N_{\rm OH}}
\Bigg(\Big[\exp\Big(-2[1+\cos(\theta_{ups})]^2\Big)\Big]
\Big[1-\tanh\left(\frac{r_{up}-r_{hs}}{t_2}\right)\Big]\Bigg)\Bigg],
\label{AH}
\end{equation}
\end{widetext}
where $u$ is the index corresponding to the neighboring clusters,
$\theta_{ups}$ is the angle between $\overrightarrow{pu}$ and
$\overrightarrow{ps}$, and $\eta$, $r_{hs}$, $t_2$ and $A_{EH}$ are
defined in Table~\ref{tab:Eparams}. Here we have again invoked a
switching function in order to avoid energy discontinuities. If a
given hydrogen atom is not involved in hydrogen bonding, then
$A_H=A_{\rm EH}$.

\begin{figure}
\includegraphics[scale=.32,angle=-90]{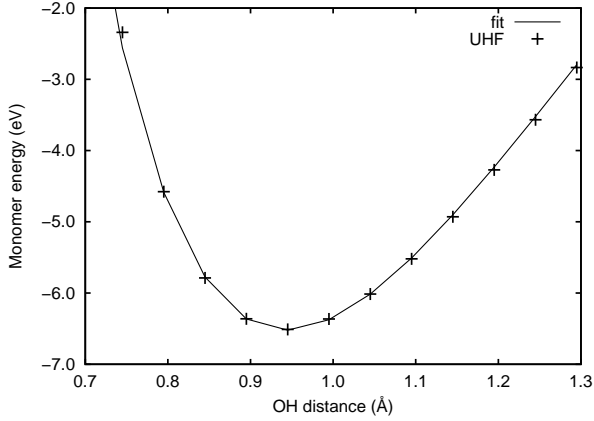}
\caption{\label{fig:Emono} UHF and predicted energies for a water
monomer for the symmetric mode at the equilibrium angle.}
\end{figure}

Figure~\ref{fig:Emono} depicts the actual (UHF calculations) and
model-predicted variation in energy of the water monomer as a
function of OH distance for the symmetric mode at the equilibrium
angle. Table~\ref{tab:monopropTable} gives a comparison of the
monomer properties as predicted by our potential, UHF calculations,
and experiment. We used a modified\cite{bfgs} BFGS routine with
analytic evaluation of gradients to determine the minimum energy
(equilibrium) geometry.

\begin{table}
\begin{ruledtabular}
\caption{\label{tab:monopropTable} Monomer equilibrium properties. }
\begin{tabular}{lrrr}
                   & Predicted\footnotemark[1]  & UHF\footnotemark[1] & Expt. \\ \hline
$R_{eq}$ (\AA)     & $0.9431$     & $0.9431$ & $0.957$\footnotemark[2]    \\
$\theta_{eq}$ ($\deg$)   &$105.47$ & $105.99$ & $104.52$\footnotemark[2]  \\
$\mu$ ($D$)   &$1.23$   & $2.19$  & $1.86$\footnotemark[3] \\
$E_{eq}$ (eV)   &$-6.51$   & $-6.52$  & ---\ \ \ \ \\
\end{tabular}
\end{ruledtabular}
\footnotetext[1]{Present work.}
\footnotetext[2]{Ref.~\onlinecite{expmono}.}
\footnotetext[3]{Ref.~\onlinecite{expmumono}.}
\end{table}

It is evident from the table as well as from Fig.~\ref{fig:Emono}
that the energetics and the minimum energy structure of the monomer
are well reproduced. However, the dipole moment of the monomer as
predicted by our potential is significantly lower than experiment.
Of course, there is no physical reason to expect the L\"owdin
charges to reproduce the dipole moments computed as proper expectation values.
Indeed, if we choose instead a definition of the atomic charge based on a physical
observable (the dipole moment),\cite{GMG98} we obtain the following
effective local (static) and nonlocal (dynamic) contributions to the
atomic charge on oxygen in the monomer: $Z^*_{\rm loc} = \mu(r)/r
|_{\rm eq} = -0.541$ and $Z^*_{\rm nl} = r\,
\partial Z^*_{\rm loc}(r)/\partial r |_{\rm eq} = -0.229$, where
$r$ refers in this case to the OH distance, and the derivative is
evaluated for the symmetric stretch mode at fixed, computed
equilibrium angle. The total Born effective charge $Z^* = -0.770$ is
given by the sum of the local and nonlocal contributions. This value
can be compared with $Z_{\rm L \ddot{o}wdin} = -0.444$. Similar
results would be expected for the dimer, where the L\"{o}wdin value
for the dipole moment is in fortuitously good agreement with
experiment.

\begin{figure}
\includegraphics[scale=.16]{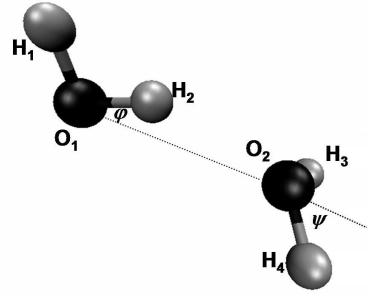}
\caption{\label{fig:dimfig} Structural parameters defining the
optimized water dimer structure.}
\end{figure}

Table~\ref{tab:dimerpropTable} contains information about the
equilibrium properties of the water dimer ({\it i.e.}, its minimum
energy configuration properties), with the geometry defined in
Fig.~\ref{fig:dimfig}. The binding energy $U$ is obtained by
subtracting the two monomer equilibrium energies from the total
energy. For comparison, we also include the relevant experimental
and UHF results. Once again, we are able to reproduce the dimer
properties reasonably well with our potential.

\begin{table}
\caption{\label{tab:dimerpropTable} Equilibrium properties of the
water dimer. All distances in \AA; angles in degrees; binding energy $U$ in
kcal/mol; $\mu$ in D.}
\begin{ruledtabular}
\begin{tabular}{lrrr}
  $$                           & Predicted\footnotemark[1] &
UHF\footnotemark[1]      & Expt.\footnotemark[2] \\ \hline
  $r_{{\rm O}_1{\rm H}_1}$     & $0.941$   & $0.948$  & ---\ \ \ \\
  $r_{{\rm O}_1{\rm H}_2}$     & $0.952$   & $0.942$  & ---\ \ \ \\
  $r_{{\rm O}_2{\rm H}_1}$     & $1.937$   & $2.038$  & ---\ \ \ \\
  $r_{{\rm O}_2{\rm H}_{4/3}}$ & $0.943$   & $0.944$  & ---\ \ \ \\
  $r_{{\rm O}_1{\rm O}_2}$     & $2.886$   & $2.98$   & $2.952$ \\
  $\angle_{{\rm H}_1{\rm O}_1{\rm H}_2}$   & $105.44$ & $105.91$ & ---\ \ \ \\
  $\angle_{{\rm H}_3{\rm O}_2{\rm H}_4}$   & $105.37$ & $106.31$ & ---\ \ \ \\
  $\angle_{{\rm O}_1{\rm H}_2{\rm O}_2}$   & $177.3$  & $179.27$ & $174.0$  \\
  $\varphi$ & $1.8$ & $-0.5$ & $0.0\pm6.0$ \\
  $\psi$    & $60.9$ & $56.7$ & $58.0\pm6.0$ \\
  $U$       & $-5.860$ & $-5.505$ & $-5.40\pm0.7$ \\
  ${\mu}$   & $2.30$   & 2.60  & $2.64$ \\
\end{tabular}
\end{ruledtabular}
\footnotetext[1] {Present work.}
\footnotetext[2]{Refs.~\onlinecite{expdim1,expdim2}.}
\end{table}

At this stage it is important to recall that the parameters in our
final energy model have been determined so as to represent the
energetics of select geometries of the water monomer, and to yield the
correct minimum energy structure of the water dimer.  No energetic
information for the remaining two model clusters used in the charge
parameterization step---${\rm H}_3{\rm O}^+$ and ${\rm OH}^-$---was
included in this {\it energy} parameterization process.
Consequently, we should not expect the model in its current form to
be able to accurately predict the energetics of ionic molecular
species. We therefore focus on assessing the predictions of the
potential for the structure and energetics of small neutral water
clusters. These results are summarized in the following section.

\section{Results}
In developing a model capable of accurately describing
water polymorphs, a basic but important requirement is the ability
to predict the correct structure and binding energies of neutral
vapor-phase water clusters.\cite{revu3} There have been numerous
computational \cite{ab1, ab2, ab3, ab4, ab5, ab6, tip4clust,
tip5clust, Fluc-q, MCDHO, POL5} and experimental studies \cite{exp1,
exp2, exp3, exp4, exp5, exp6, exp7, exp8, exp9} examining various
water clusters. It has been shown that small neutral water clusters
have 2D cyclic structures, where each molecule serves as both an
acceptor and a donor, while the larger clusters have 3D structures.
This crossover is seen for the water hexamer and
larger clusters, where the 3D structures are energetically favored.
Some of the popular water potentials (MCDHO, TIP5P, TIP4P, POL5,
Dang and Chang (DC)) have been used to study small water clusters
with varying degrees of success. In the following, we compare our
results with these potentials, as well as experiments and quantum
calculations. We pay particular attention to the different
structures of the hexamer.

\subsection{Small water clusters: trimer-pentamer}
\begin{figure}
\includegraphics[scale=.5]{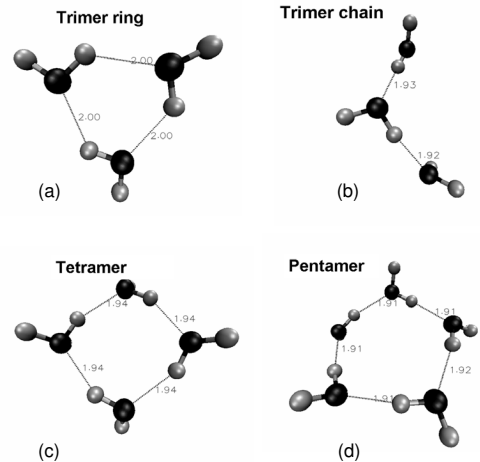}
\caption{Equilibrium geometries of ${\rm H}_2{\rm O}_n$, $n$=3-5;
interatomic distances in \AA .} \label{fig:35clust}
\end{figure}

Early spectroscopic studies\cite{exp9} predicted the open chain
conformation to be the most stable structure for the trimer.
Subsequent work has suggested otherwise,\cite{exp2, ab4} and the
cyclic trimer with $C_1$ symmetry has been shown to be the more
stable structure. Using the modified BFGS routine to perform the
energy minimization, we found the ring structure to be slightly more
stable than the open chain conformation with the difference in
energy being $0.861$ kcal/mol (this lies within the margin of error
for our potential fit, 0.02 eV/molecule $\times$ 3 = 1.38 kcal/mol.)
Next, we obtained the energies and optimized geometries of the
predicted ground-state structure of the tetramer and pentamer. The
$S_4$ cyclic tetramer structure has been shown to be most
energetically favored for the water tetramer.\cite{exp8} In this
structure (Fig.~\ref{fig:35clust}(c)), there are alternating
hydrogen atoms above and below the plane of the tetramer ring. A
puckered cyclic ring (Fig.~\ref{fig:35clust}(d)) has been predicted
to be the most stable pentamer structure by both {\it ab initio}
calculations \cite{ab4} and experiment.\cite{exp4}

Table~\ref{tab:tri2penprop} gives the properties of the three
clusters; for comparison, along the lines of Stern {\it et
al.},\cite{POL5} we present results of select potentials along with
\textit{ab initio} calculations and experiment.  The notation
$\langle\cdots\rangle$ in
Tables~\ref{tab:tri2penprop}--\ref{tab:10prop} reflects the fact
that all quoted distances are averaged over the cluster structure.
Though our model predicts the correct structure and energetics, the
net dipole moment $\mu$ is once again smaller than that computed via
other models as well as experiment, as expected based on our
previous discussion. Scaling $\mu$ by a factor equal to the ratio of
the experimental monomer dipole moment and our model's monomer
dipole moment ($\mu_{norm}$= 0.6613) yields values that are more
realistic; these are the values reported in
Table~\ref{tab:tri2penprop}.

\begin{table*}
\caption{\label{tab:tri2penprop} Equilibrium properties of water
clusters ($n$=3--5). All distances in \AA, angles in degrees,
energies in kcal/mol, charge in $e$, and dipole moment $\mu$ in
debye (D). 6-31G** HF energies are given in parentheses along with
select \textit{ab initio} values.}
\begin{ruledtabular}
\begin{tabular}{crrrrrrrrr}
  $$ & Predicted & POL5/TZ\footnotemark[1] & POL5/QZ\footnotemark[1] & TIP4P/FQ\footnotemark[2]
  & TIP5P\footnotemark[3] & MCDHO\footnotemark[4] & $\textit{ab initio}$ & Expt. \\ \hline
Trimer-- Cyclic & \\
  $U$ &  $-13.743$  & $-13.416$ &$-13.453$& $-12.576$& $-14.992$ & $-13.982$& $-15.9$\footnotemark[5] $(-17.10)$& $$  \\
  $\langle r_{\rm OO} \rangle$  & $2.712$  & $2.901$ & $2.893$ & $2.912$ & $2.770$ & $2.911$ & $2.782$\footnotemark[5] &
  $2.960$\footnotemark[8] & \\
  ${\mu}$  &$0.673$   & $1.205$ & $1.205$ & $0.417$  & $1.074$ & $1.114$ & $1.071$\footnotemark[6] & $$ \\
   $\langle \angle_{\rm HOH} \rangle$& $105.26$ \\
  $\langle r_{\rm OH} \rangle$ & $0.949$ \\
  $\langle q_{\rm H} \rangle$ &  $0.221$ \\ \\

Tetramer-- Cyclic \\
  $U$ &$-27.308$  & $-25.529$ &$-25.665$& $-23.641$& $-28.431$ & $-27.581$& $-23.8$\footnotemark[7] $ (-29.10)$& $$&  \\
  $\langle r_{\rm OO}\rangle$  & $2.826$  & $2.769$ & $2.759$& $2.809$ & $2.673$ & $2.806$ & $2.743$\footnotemark[7] &
  $2.79$\footnotemark[8] \\
  ${\mu}$  &$0.015$   & $0.000$ & $0.000$ & $0.000$  & $0.000$ & $0.024$ & $0.000$\footnotemark[6] & $$ \\
   $\langle \angle_{\rm HOH}\rangle$& $105.27$ \\
  $\langle r_{\rm OH}\rangle $ & $0.948$ \\
  $\langle q_{\rm H} \rangle$ &  $0.220$ \\ \\

Pentamer-- Cyclic \\
$U$   &  $-36.027$  & $-34.111$ &$-34.427$& $-32.954$& $-38.122$ & $-35.229$& $-33.34$\footnotemark[7] $ (-37.70)$& $$ \\
  $\langle r_{\rm OO}\rangle$  & $2.850$  & $2.742$ & $2.726$ & $2.773$ & $2.657$ & $2.753$ & $2.867 $\footnotemark[7] &
  $2.760$\footnotemark[8] \\
  ${\mu}$ & $0.634$   & $1.190$ & $1.191$ & $0.401$  & $1.219$ & $0.992$ & $0.927$\footnotemark[6] & $$
  &\\
  $\langle \angle_{\rm HOH}\rangle$& $105.26$ \\
  $\langle r_{\rm OH}\rangle$ & $0.948$ \\
  $\langle q_{\rm H}\rangle$ &  $0.220$ \\
\end{tabular}
\end{ruledtabular}
\footnotetext[1]{Ref.~\onlinecite{POL5}.}
\footnotetext[2]{Refs.~\onlinecite{Fluc-q,POL5}.}
\footnotetext[3]{Ref.~\onlinecite{TIP5P}.}
\footnotetext[4]{Ref.~\onlinecite{MCDHO}.}
\footnotetext[5]{Ref.~\onlinecite{ab5}.}
\footnotetext[6]{Refs.~\onlinecite{ab7,ab8}.}
\footnotetext[7]{Ref.~\onlinecite{ab4}.}
\footnotetext[8]{Refs.~\onlinecite{exp1,exp2} (trimer);
\onlinecite{exp7,exp8} (tetramer);\ \onlinecite{exp3,exp4}
(pentamer).}
\end{table*}

\subsection{Water hexamer}
\begin{figure}
\includegraphics[scale=.5]{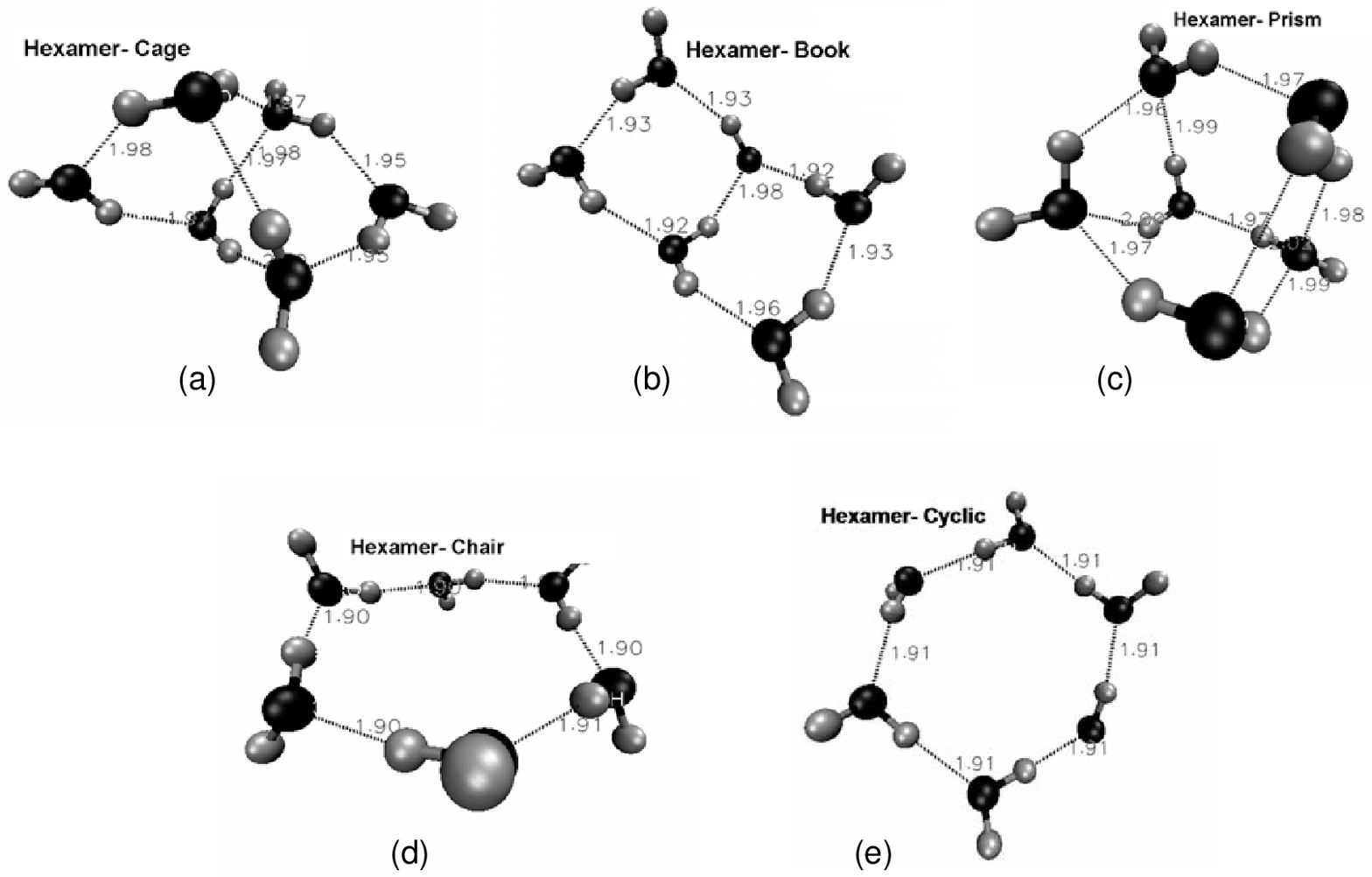}
\caption{Equilibrium geometries of water hexamers; interatomic
distances in \AA .} \label{fig:6clust}
\end{figure}

For the water hexamer, it has now been established that there are a
number of different local minima structures that are energetically
very comparable. IR spectroscopic experiments on gas-phase clusters by Paul {\it et al}.
\cite{paul} and Liu {\it et al}. \cite{exp5, exp6} indicate that the
caged hexamer structure is the most stable, while \textit{ab initio}
calculations have revealed that the cage, prism and book structures
are almost degenerate, with the stability sequence depending on the
inclusion of zero-point energy differences.\cite{hex1, hex2, hex3,
hex4, hex5, hex6, hex7}
Further, Tissandier {\it et al}.\cite{hexref} have used a
topological enumeration technique in conjunction with semi-empirical
PM3 methods to predict the global minimum energy structures. Here we
examine the cyclic, cage, prism, chair and book structures; the
results are provided in Tables~\ref{tab:hexprop1}
and~\ref{tab:hexprop2}. The prism, book and the cage structures are
the most stable and are energetically nearly degenerate, while the
cyclic and chair are clearly metastable structures at 0 K. The
computed dipole moment $\mu$ has been scaled by $\mu_{norm}$.
Fig.~\ref{fig:6clust} shows the various water hexamers as obtained
from our model. The results clearly indicate that our model is
capable of describing the experimentally determined structures and
relative energetics of the water hexamers.

\begin{table*}
\begin{ruledtabular}
\caption{\label{tab:hexprop1} Equilibrium properties of water
hexamers I. All distances in \AA, angles in degrees, energy in
kcal/mol, charge in $e$, and dipole moment $\mu$ in D.\ 6-31G** HF
energies are given in parentheses along with select \textit{ab
initio} values.}
\begin{tabular}{crrrrrrrrrr}
  $$ & Predicted  & POL5/TZ\footnotemark[1] & POL5/QZ\footnotemark[1] & TIP4P/FQ\footnotemark[2] &
  TIP5P\footnotemark[3] & MCDHO\footnotemark[4] & DC\footnotemark[5] & $\textit{ab initio}$ &
  Expt. \\ \hline

Hexamer-- Cage \\
  $U$   &  $-46.497$ &$-41.783$  & $-39.297$ &$-45.388$& $-45.388$& $-43.690$ & $-40.76 $& $-45.03$\footnotemark[6] $(-48.60)$& $$ \\
  $\langle r_{\rm OO}\rangle$  & $2.801$ & $2.783$  & $2.755$ & $2.863$ & $2.746$ & $2.888$ & $$& $2.807$\footnotemark[6] &
  $2.820 $\footnotemark[8] \\
  ${\mu}$  &$2.120$   & $2.442$ & $2.454$ & $1.788$  & $2.178$ & $2.034$ & $$ & $2.05$\footnotemark[7] &
  $1.904$\footnotemark[8] \\
   $\langle \angle_{\rm HOH}\rangle$& $105.14$ &\\
  $\langle r_{\rm OH}\rangle $ & $0.950$ \\
  $\langle q_{\rm H}\rangle$ &  $0.220$ \\ \\

Hexamer-- Book \\
  $U$   &  $-46.492$ &$-42.464$  & $-42.771$ &$-40.152$& $-46.680$& $-43.977$ & $-40.38$& $-44.74$\footnotemark[6] & $$ \\
  $\langle r_{\rm OO}\rangle$  & $2.788$ & $2.777$  & $2.777$ & $2.815$ & $2.688$ & $2.809$ & $$& $2.766$\footnotemark[6] &  $$ \\
  ${\mu}$  &$2.410$   & $2.449$ & $2.430$ & $2.006$  & $2.445$ & $$ & $$ & $$ &  $$ \\
   $\langle \angle_{\rm HOH}\rangle$& $105.22$ \\
  $\langle r_{\rm OH}\rangle$ & $0.949$ \\
  $\langle q_{\rm H}\rangle$ &  $0.220$ \\ \\

Hexamer-- Prism & \\
  $U$   &  $-46.465$ &$-41.847$  & $-42.135$ &$-39.304$& $-45.805$& $-44.192$ & $-40.97$& $-45.12$\footnotemark[6]
  $(-49.60)$& $$ \\
  $\langle r_{\rm OO}\rangle$  & $2.757$ & $2.792$  & $2.782$ & $2.819$ & $2.773$ & $2.892$ & $$& $2.840$\footnotemark[6] &  $$ \\
  ${\mu}$  &$2.974$   & $2.905$ & $2.931$ & $3.254$  & $2.692$ & $2.627$ & $$ & $2.701$\footnotemark[7] &  $$ \\
   $\langle \angle_{\rm HOH}\rangle $& $104.85$ \\
  $\langle r_{\rm OH}\rangle$ & $0.951$ \\
  $\langle q_{\rm H}\rangle$ &  $0.219$ \\
\end{tabular}
\end{ruledtabular}
\footnotetext[1]{Ref.~\onlinecite{POL5}.}
\footnotetext[2]{Ref.~\onlinecite{Fluc-q,POL5}.}
\footnotetext[3]{Ref.~\onlinecite{TIP5P}.}
\footnotetext[4]{Ref.~\onlinecite{MCDHO}.}
\footnotetext[5]{Ref.~\onlinecite{DC1}.}
\footnotetext[6]{Ref.~\onlinecite{ab3}.}
\footnotetext[7]{Refs.~\onlinecite{ab7,ab8}.}
\footnotetext[8]{Refs.~\onlinecite{exp5,exp6}.}
\end{table*}

\begin{table*}
\caption{\label{tab:hexprop2} Equilibrium properties of water
cluster hexamers II. All distances in \AA, angles in degrees, energy
in kcal/mol, charge in $e$, and dipole moment $\mu$ in D.\ 6-31G**
HF energies are given in parentheses along with select \textit{ab
initio} values.  Footnotes as in Table~\ref{tab:hexprop1}.}
\begin{ruledtabular}
\begin{tabular}{crrrrrrrrrr}
$$ & Predicted    & POL5/TZ\footnotemark[1] & POL5/QZ\footnotemark[1] & TIP4P/FQ\footnotemark[2] &
   TIP5P\footnotemark[3] & MCDHO\footnotemark[4] &  DC\footnotemark[5] & $\textit{ab initio}$ & Expt. \\
\hline

Hexamer-- Chair \\
  $U$   &  $-44.073$ \\
  $\langle r_{\rm OO}\rangle$  & $2.846$ \\
  ${\mu}$  &$0.011$  \\
   $\langle\angle_{\rm HOH}\rangle$& $105.25$ \\
  $\langle r_{\rm OH}\rangle$ & $0.948$ \\
  $\langle q_{\rm H}\rangle$ &  $0.220$ \\ \\

Hexamer-- Cyclic \\
  $U$   &  $-43.919$ &$-41.875$  & $-42.224$ &$-41.368$& $-47.309$& $-44.264$ & $-39.34$& $-43.88$\footnotemark[6] & $$ \\
  $\langle r_{OO}\rangle$  & $2.849$ & $2.737$  & $2.720$ & $2.756$ & $2.654$ & $2.731$ & $$& $2.714$\footnotemark[6] &
  $2.756$\footnotemark[7] \\
  ${\mu}$  &$0.150$   & $0.017$ & $0.003$ & $0.000$  & $0.000$ & $0.134$ & $$ & $0.000$ &  $$ \\
   $\langle \angle_{HOH}\rangle$& $105.25$ \\
  $\langle r_{OH}\rangle$ & $0.948$ \\
  $\langle q_{\rm H}\rangle$ &  $0.220$ \\ \\

\end{tabular}
\end{ruledtabular}
\end{table*}

\subsection{Beyond the Hexamer}
The experimental energetics and the structures of water clusters
with six or fewer molecules have been well documented.\cite{expmono,
expmumono, expdim1, expdim2, exp1, exp2, exp3, exp4, exp5, exp6,
exp7, exp8} This is not true for larger water clusters ($n$ $\ge$
10), and information about such clusters is available mainly via
classical potentials and quantum calculations. Hence we compare our
results only with other computational studies.\cite{ab2,tip4clust,
tip5clust} Maheshwary {\it et al}. \cite{ab2} have examined the
structure and stability of water clusters (up to twenty-molecule
clusters) using Hartree Fock as well as DFT (B3LYP) calculations
with 6-31G** and 6-31++G** basis sets; calculations using TIP4P
\cite{tip4clust} and TIP5P \cite{tip5clust} potentials have also
been performed for these clusters.

\begin{figure}
\includegraphics[scale=.5]{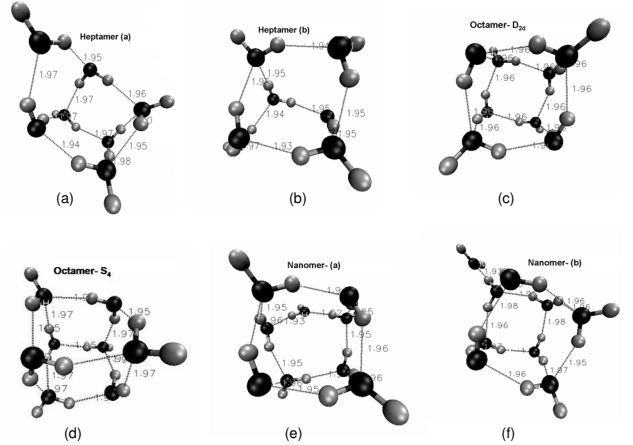}
\caption{Equilibrium geometries of ${\rm H}_2{\rm O}_n$, $n$=7-9;
interatomic distances in \AA .} \label{fig:79cluster}
\end{figure}

The experimentally-determined\cite{sep1} and theoretically predicted
\cite{ab2} stable heptamer conformer is a cuboid structure with a
missing corner, labeled Heptamer~($a$) in Fig.~\ref{fig:79cluster}.
This is also the lowest energy geometry as predicted by our
potential, with an unscaled dipole moment of 1.20~D. In addition, we
observe another structure (Heptamer~($b$) in
Fig.~\ref{fig:79cluster}) to be approximately $1$ kcal/mol higher in
energy. This structure has a high dipole moment (3.94~D), and nine
hydrogen bonds, in contrast to the ten found in the more stable
conformer. Note that the dipole moments reported in
Table~\ref{tab:79prop} for the large clusters are as obtained and
have not been rescaled, since no experimental data is available for
comparison.  (We also would expect the deviation between theory and
experiment resulting from our specific choice of atom-in-molecule
charge definition to ``wash out" for the larger clusters.) The same
ordering in the energies and dipole moments is seen in the work of
Maheshwary {\it et al.}\cite{ab2}

The most stable state of the water octamer in our work is cubic with
$D_{2d}$ symmetry. The next most stable octamer structure is another
cubic structure with $S4$ symmetry. We observe a difference of
almost 1.4 kcal/mol in the relative energies of the two structures;
Maheshwary {\it et al.}\cite{ab2} predict the two structures to be
nearly isoenergetic. The dipole moment is zero for both structures,
with each structure characterized by twelve hydrogen bonds. These
structures are shown in Fig.~\ref{fig:79cluster}.

The global minimum water nanomer structure can be described in terms
of a pentamer and a tetramer ring connected by hydrogen bonds
(Nanomer~($a$) in Fig.~\ref{fig:79cluster}). This structure is seen
by experimental studies of Buck {\it et al}. \cite{buck} as well as
computational studies by Maheshwary {\it et al}. \cite{ab2} and Dang
and Chang,\cite{DC1} and is characterized by thirteen hydrogen
bonds. Our potential also predicts this structure to be the most
stable. Another stationary point on the nanomer energy surface is
the structure Nanomer~($b$) as shown in Fig.~\ref{fig:79cluster}.
This structure contains 13 hydrogen bonds, and can be described as a
octamer cube plus a monomer coordinated to a corner of the cube via
a hydrogen bond.

\begin{table}
\caption{\label{tab:79prop} Equilibrium properties of water clusters
for $n$=7-9. All distances in \AA, angles in degrees, energy in
kcal/mol, charge in $e$, and dipole moment $\mu$ in $D$.}
\begin{ruledtabular}
\begin{tabular}{crrrrr}
$$ & Predicted    & TIP4P\footnotemark[1] &  TIP5P\footnotemark[2] &  \textit{ab initio}\footnotemark[3] \\ \hline

Heptamer ($a$) \\
  $U$   &  $-58.259$ &$-58.271$& $-57.910$&  $-60.53$ \\
  $\langle r_{\rm OO}\rangle$  & $2.802$ & $2.762$  & $2.738$  & $2.884$   \\
  ${\mu}$  &$1.20$   & $$ & $$ &   $1.35$ \\
   $\langle \angle_{\rm HOH}\rangle $& $105.09$ & $$ &  $$& $106.26$ \\
  $\langle r_{\rm OH} \rangle$ & $0.951$ & $$ & $$&  $0.950$ \\
  $\langle q_{\rm H}\rangle $ &  $0.219$ \\ \\

Octamer $D_{2d}$ \\
   $U$   &  $-74.325$ &$-73.090$& $-72.535$ & $-76.01$ \\
  $\langle r_{\rm OO}\rangle$  & $2.822$ & $2.746$  & $2.712$ &  $2.877$ \\
  ${\mu}$  &$0.00$   & $$ & $$ &  $0.00$ \\
   $\langle \angle_{\rm HOH} \rangle$& $105.13$ &  $$& $$& $106.482$ \\
  $\langle r_{\rm OH} \rangle$ & $0.951$ & $$ & $$& $0.951$ \\
  $\langle q_{\rm H}\rangle$ &  $0.219$ \\ \\

Nanomer ($a$) \\
  $U$   &  $-84.124$ &$-82.401$& $-83.622$& $-85.05$    \\
  $\langle r_{OO}\rangle$  & $2.842$ & $2.741$  & $2.696$  & $2.869$  \\
  ${\mu}$  &$0.99$   & $$ & $$ &  $1.69$ \\
   $\langle \angle_{\rm HOH}\rangle $& $105.16$ & $$ & $$&  $106.39$ \\
  $\langle r_{\rm OH}\rangle $ & $0.950$ & $$ & $$&  $0.950$ \\
  $\langle q_{\rm H}\rangle$ &  $0.219$ \\ \\
\end{tabular}
\end{ruledtabular}
\footnotetext[1]{Ref.~\onlinecite{tip4clust}.}
\footnotetext[2]{Ref.~\onlinecite{tip5clust}.}
\footnotetext[3]{Ref.~\onlinecite{ab2}.}
\end{table}

Locating the global energy minimum for larger clusters ($n \geq 10$)
is a difficult task since the flat potential energy surface gives
rise to many possible geometries with comparable energies. We have
therefore used the geometries predicted by Maheshwary {\it et
al.},\cite{ab2} TIP5P,\cite{tip5clust} and TIP4P\cite{tip4clust}
(available online at the Cambridge cluster database website \cite{camb}) as
starting configurations for our energy minimization calculations.
The energies of our resulting energy-minimized structures for
$n=10-20$ (Table~\ref{tab:10prop}), agree reasonably well with the
calculations of Maheshwary {\it et al}. Rather than providing the
geometries of all the above clusters, we have listed their important
properties in Table~\ref{tab:10prop}; the table also indicates the
initial structure that yields the minimum energy geometry when we
perform our minimization.

\begin{figure}
\rotatebox{-90}{
\includegraphics[scale=0.32,angle=0]{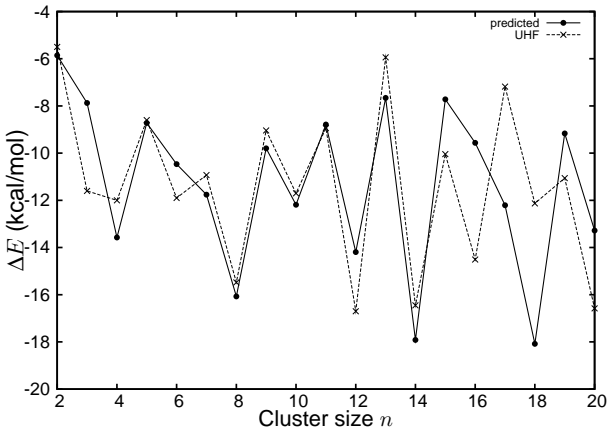}}
\caption{Incremental interaction energies of water clusters, $\Delta
E$ = $E_{n+1} - E_n - E_1$, as a function of cluster size $n$.  Note that
$\Delta E$ = $U_{n+1} - U_n$, where $U_n$ is the binding energy for a
cluster of size $n$.  UHF
results from Ref.~[\onlinecite{ab2}].} \label{fig:Ediff}
\end{figure}

A summary comparison of our model-predicted results with the {\it ab
initio} calculations of Maheshwary {\it et al}. is given in
Fig.~\ref{fig:Ediff}.  Some deviations from the {\it ab initio}
results occur at $n=3$, 6, and 16.  In particular, as shown in
Table~\ref{tab:tri2penprop}, we underestimate the binding energy of
the trimer, leading to the deviation in estimation of the
incremental interaction energy at $n=3$.  Although the model is able
to predict the correct ordering of the binding energies of the
various hexamers and heptamer, it is unable to capture the small
difference in incremental interaction energy between $n=6$ and
$n=7$. However, we do largely reproduce the alternation in stability
of the cluster depending on whether $n$ is odd or even---in
particular, the enhanced stability of even $n$-mers relative to odd
$n$-mers.

\begin{table*}
\caption{\label{tab:10prop} Predicted equilibrium cluster-averaged
properties of water clusters for $n\geq 10$. All distances in \AA,
angles in degrees, energy in kcal/mol, charge in $e$, and dipole
moment $\mu$ in $D$.  ``Geometry" refers to the starting geometry
for the energy minimization, as described in the text.}
\begin{ruledtabular}
\begin{tabular}{ccccccccc}
  $n$ & $U$    & $\langle r_{\rm OO} \rangle$ & ${\mu}$& $\langle \angle_{\rm HOH}\rangle$ &
  $\langle r_{\rm OH}\rangle$ & $\langle q_{\rm H}\rangle$ &  Geometry \\ \hline

10
     &  $-96.311$&  $2.841$& $1.81$ &
  $105.12$ &
   $0.951$ &
    $0.219$ &
   TIP5P \\

11
     &  $-105.098$& $2.849$& $2.54$ &
  $104.97$ &
   $0.952$ &
    $0.219$ &
   TIP5P \\

12
     &  $-119.291$& $2.828$& $0.00$ &
   $104.87$ &
   $0.953$ &
    $0.219$ &
   Ref.~[\onlinecite{ab2}] \\

13
     &  $-126.951$& $2.835$& $1.65$ &
  $105.11$ &
  $0.951$ &
  $0.219$ &
  TIP4P \\

14
     &  $-144.868$& $2.835$& $1.86$ &
  $105.01$ &
  $0.952$ &
  $0.218$ &
  TIP4P \\

15
     &  $-152.589$& $2.850$& $1.98$ &
  $105.12$ &
  $0.951$ &
  $0.218$ &
  TIP5P \\

16
     &  $-162.153$& $2.829$& $0.00$ &
  $104.73$ &
  $0.954$ &
  $0.219$ &
  TIP4P \\

17
     &  $-174.362$& $2.848$& $3.03$ &
  $105.02$ &
  $0.952$ &
  $0.219$ &
  TIP5P \\

18
     &  $-192.442$& $2.836$& $1.85$ &
  $104.93$ &
  $0.953$ &
  $0.218$ &
  TIP4P \\

19
     &  $-201.608$& $2.842$& $2.97$ &
  $105.03$ &
  $0.952$ &
  $0.218$ &
  TIP5P \\

20
     &  $-214.889$& $2.833$& $0.17$ &
  $104.93$ &
  $0.953$ &
  $0.218$ &
  TIP4P \\
\end{tabular}
\end{ruledtabular}
\end{table*}

\section{Discussion and Conclusions}
We have presented a new dynamical CT-EAM potential for modeling
water and its polymorphs at the atomic level. We have based our
parameterization on \textit{ab initio} data; in particular, atomic
charge fluctuations have been modelled with reference to the local
chemical environment, using L\"owdin population analysis to
represent atomic charges. Depending on its immediate coordination
environment, each atom is assigned to a cluster, with this
identification being crucial to our formulations.
Cluster identification is effected via a radial
cutoff, and total cluster charge is based on the size of the cluster and
relative positions of neighboring clusters. This charge is
in turn partitioned among the constituent atoms.

Our technique is sufficiently flexible to account for very different
charge states of clusters and individual atoms.  The radial cutoff
chosen to define our clusters, $r_{cut} = 1.5$ \AA , is
significantly larger than the O$-$H equilibrium distance in the
monomer and dimer ($\sim 0.95$ \AA .)  Consequently, the model is
easily capable of describing non-perturbative charge transfer.

We note that a number of important effects have been
omitted in this initial implementation.  This was done in order to focus
attention on the physics of the charge-transfer EAM model framework itself,
rather than the refinement of a model water potential {\it per se.}
For example, the current parameterization does not yet impose the
correct asymptotic dissociation behavior on cluster subsystems,
which we have argued is critical to a proper description of reactive
dynamics.\cite{vat06a,PPLB} A related issue concerns the omission of
several ionic species, believed to be important in defining the
hydrogen network in water, from our parameterizations: these include
the Zundel (${\rm H}_5{\rm O}_2^+$) and Eigen (${\rm H}_9{\rm
O}_4^+$) cations.\cite{MEVB} A final important simplification
concerns the use of atom-in-molecule charges as proxies for the
shape function modeling of the AIM charge-density distributions. It
is clear that further work taking account of these various factors
will be necessary in order to successfully study complex kinetic
processes such as those involved in ion solvation, enzyme catalysis,
and proton transport. This work is presently underway. Additionally,
it should be noted that our approach does not incorporate a quantum
mechanical treatment of the actual electron or proton transfer
processes.\cite{PCET}

Notwithstanding the simplicity of this initial model, in tests on
small water clusters, our results agree very well with experimental
and {\it ab initio} data. Importantly, our model captures the
transition from planar ring pentamer structures to three-dimensional
complex hexamer structures, an essential structural test for any
successful water potential. In this context, it is worth noting that
an environment dependent dynamic charge potential motivated by the
present work has also been developed recently for silica. This
potential successfully matches {\it ab initio} results in its
ability to predict the ground-state energy, geometry and failure
mechanisms of silica clusters.\cite{KRM06}

More generally, this work represents a successful application of
many-body embedded atom concepts to the modeling of a highly
polarizable {\it molecular} system, and thus a significant departure from
traditional approaches to developing water potentials.
It is remarkable that even this relatively simple implementation of
CT-EAM reproduces cluster structures and energetics consistent with
the best previous potentials, while providing a theoretical roadmap
for implementing true charge-transfer dynamics.  This ability of an
embedded-atom approach---originally designed for describing
many-body effects in bulk {\it fcc} metals---to model the structure
of a molecular system can be understood as a direct consequence the
CT-EAM framework's underlying density functional construction.
DFT, with its emphasis on electron densities
as the fundamental variables of the theory, acts as a multiscale
mechanism for incorporating quantum mechanical bonding effects
and excitations within a nominally classical potential.

We believe that the unique combination of features described here
will ultimately enable CT-EAM potentials to successfully capture
many-body and electrostatic effects, in both static and dynamic
contexts, for a wide variety of biophysical and materials systems,
including nanoscale systems possessing mixed molecular and bulk
features. Future applications
will include studies of the crystalline polymorphs of water and the
thermodynamic and structural properties of the liquid, as well as
investigations of dynamical processes such as ion solvation and
proton transport.

\section{Acknowledgements}
We thank Dr.~Keith Runge (University of Florida) for many useful
insights, Dr.~Andy~Pineda (University of New Mexico) for assistance
with {\tt GAMESS}, and the UNM Center for High Performance Computing
for computational resources. This work was supported by NSF Grant
No.~CHE-0304710. S.R.A. gratefully acknowledges support from NSF
Grant No.~DMR-9520371 during the early stages of this research. The
work of S.M.V. was performed at Los Alamos National Laboratory under
the auspices of the U.S. Department of Energy, under contract
No.~DE-AC52-06NA25396. One of the authors (K.M.) would like to
thank Prof.~Sam Trickey and the University of Florida
Quantum Theory Project for postdoctoral support under NSF ITR award DMR-0325553,
where portions of this work were completed.


\begin{thebibliography}{100}
\bibitem{fn1} The L\"{o}wdin atomic charge $q_A$ for an atom $A$ is defined as
$q_A$ = $Z_A - \sum_{\mu \in A} (S^{1/2}PS^{1/2})_{\mu \mu}$, where
$Z_A$ is the atomic charge, $S$ is the matrix of basis function
overlap integrals, and $P$ is the density matrix expressed in terms
of orbital basis-set expansion coefficients.\cite{Lowdin}
\end{thebibliography}
\end{document}